\documentclass[a4paper, amsfonts, amssymb, amsmath, reprint, showkeys, nofootinbib, twoside, onecolumn,notitlepage]{revtex4-1}

\def\mnras{{MNRAS}}             
\def\prd{{Phys.~Rev.~D}}        
\def\pre{{Phys.~Rev.~E}}        
\def\prl{{Phys.~Rev.~Lett.}}    
\usepackage{amsmath,amstext}
\usepackage[T1]{fontenc}
\usepackage{amssymb}
\input{epsf}
\usepackage{graphicx}
\usepackage{ae,aecompl}

\usepackage{hyperref}
\usepackage{amsmath}
\usepackage{amssymb}
\usepackage{mathtools}
\usepackage{bm}
\usepackage{cleveref}
\usepackage{tensor}
\usepackage{braket}
\usepackage{enumitem}
\usepackage{mhchem}
\usepackage{cancel}

\usepackage{color}

\newcommand{\spc}{\quad \quad \quad}

\def\be{\begin{equation}}
\def\ee{\end{equation}}
\def\beq{\begin{eqnarray}}
\def\eeq{\end{eqnarray}}

\begin{document}
\title{Is Relativistic Hydrodynamics always Symmetric-Hyperbolic in the Linear Regime?}
\author{L.~Gavassino}
\affiliation{Department of Mathematics, Vanderbilt University, Nashville, TN, USA}

\begin{abstract}
Close to equilibrium, the kinetic coefficients of a thermodynamic system must satisfy a set of symmetry conditions, which follow from the Onsager-Casimir principle. Here, we show that, if a system of hydrodynamic equations is analysed from the perspective of the Onsager-Casimir principle, then it is possible to impose very strong symmetry conditions also on the principal part of such equations (the part with highest derivatives). In particular, we find that, in the absence of macroscopic magnetic fields and spins, relativistic hydrodynamics should always be symmetric-hyperbolic, when linearised about equilibrium. We use these results to prove that Carter's multifluid theory and the Israel-Stewart theory in the pressure frame are both symmetric-hyperbolic in the linear regime. Connections with the GENERIC formalism are also explored.
\end{abstract}

\maketitle

\section{Introduction} 

Symmetric-hyperbolicity is a highly desirable property for a hydrodynamic theory \cite{CourantHilbert2_book,GerochPartial1996,
Frittelli1996}. If a quasilinear system of partial differential equations is symmetric-hyperbolic, then we know, under standard regularity assumptions \cite{Kato2009}, that the solutions to the initial-value problem exist, are unique, and depend continuously on the initial data (at least for some finite time). Unfortunately, assessing whether a hydrodynamic theory admits a symmetric formulation in the fully non-linear regime is hard. So far, we only know of few theories that admit such a formulation: divergence-type theories \cite{GerochLindblom1990} and more generally any theory that is consistent with the principles of Rational Extended Thermodynamics \cite{Muller_book,Ruggeri1981,RuggeriMagneto1981}, Geroch-Lindblom theories \cite{Geroch_Lindblom_1991_causal,
LindblomRelaxation1996,GavassinoNonHydro2022} (which are symmetric by construction), and the Israel-Stewart theory with \textit{only} bulk viscosity \cite{Causality_bulk}. On the other hand, if we restrict our attention to linear deviations from global thermodynamic equilibrium, the situation improves considerably, as the full Israel-Stewart theory acquires a symmetric form, both in the Eckart \cite{Hishcock1983} and in the Landau frame \cite{OlsonLifsh1990}. As a consequence, those theories that (close to equilibrium) become indistinguishable from the Israel-Stewart theory \cite{noto_rel,PriouCOMPAR1991,Lopez09,Otting1_1998,
Denicol2012Boltzmann,BulkGavassino,
GavassinoRadiazione,Stricker2019,GavassinoGENERIC2022} are also symmetric-hyperbolic, when linearised. This covers most of the relativistic hydrodynamic theories whose entropy current has strictly non-negative divergence ($\partial_\mu s^\mu \geq 0$).

Our goal here is to understand if hydrodynamics \textit{should be} symmetric-hyperbolic in the linear regime. Do we have any physical reason to believe that the principal part of the hydrodynamic equations must be symmetric? There is only one universal principle that may be used to impose such a strong symmetry condition on the kinetic coefficients of a system: the Onsager-Casimir principle \cite{Onsager_1931,Onsager_Casimir,landau5,Meixner1963,
Kubo1957,Pavelka2014}. 
In a nutshell, the Onsager-Casimir principle states that the dynamical coupling between degrees of freedom which acquire the same phase under time reversal (``even-even'' or ``odd-odd'' case) is symmetric, while the dynamical coupling between degrees of freedom which acquire opposite phases under time reversal (``even-odd'' case) is antisymmetric.
Let's see what conclusions can be drawn from it.

We adopt the metric signature $(-,+,+,+)$ and work with natural units $c=k_B=1$. We also adopt the following notation: $\mu,\nu,\rho$ are space-time indices, while $A,B,C$ are field multi-indices, defined below. The indices $j,k,l$ are purely spacelike indices: they run from 1 to 3. Einstein's summation convention applies to all kinds of indices. 

\section{Intuitive connection between symmetric hyperbolicity and Onsager symmetry}

In this section, we provide a brief introduction to the Onsager-Casimir principle and to the notion of symmetric hyperbolicity. Then, we present a simple toy model, which should convince the reader that there is, indeed, a deep connection between these two seemingly unrelated concepts, the former arising from the principle of detailed balance \cite{Onsager_1931}, and the latter arising from the requirement that the equations should be solvable \cite{Kato2009}.

\subsection{Brief overview of the Onsager-Casimir principle}

The original formulation of the Onsager-Casimir (OC) principle, due to \citet{Onsager_1931}, is best illustrated if one considers a chemistry-related example. Suppose that $\delta\mu^1$ and $\delta\mu^2$ are the linear perturbations to the chemical potentials of two non-conserved particle species (e.g. photons and neutral pions) in a homogeneous fluid. In thermodynamic equilibrium, one has $\delta\mu^1=\delta\mu^2=0$. Then, since the entropy is maximised at equilibrium, we can expand the entropy perturbation $\Delta S=S-S_{\text{eq}}$ to second order in the fluctuations:
\begin{equation}\label{Susz}
\Delta S= -\dfrac{1}{2} 
\begin{bmatrix}
\delta\mu^1 &  \delta\mu^2
\end{bmatrix}
\begin{bmatrix}
S_{11} & S_{12} \\
S_{21} & S_{22}  \\
\end{bmatrix}
\begin{bmatrix}
\delta \mu^1 \\
\delta \mu^2 \\
\end{bmatrix} ,
\end{equation} 
with $S_{12}=S_{21}$. Now, for sufficiently small deviations from equilibrium, we can also assume that the functions $\delta\mu^1(t)$ and $\delta\mu^2(t)$ obey some coupled linear differential equations of the form $ \delta\dot{\mu}^A = \mathcal{K}^A_B  \delta\mu^A$, for some kinetic matrix $\mathcal{K}^A_B$. However, since the $2 \times 2$ matrix in \eqref{Susz} is invertible (being positive definite), we can always rewrite the  dynamical system $ \delta \dot{\mu}^A = \mathcal{K}^A_B \delta\mu^B$ in the equivalent form below:
\begin{equation}
\begin{bmatrix}
S_{11} & S_{12} \\
S_{21} & S_{22}  \\
\end{bmatrix}
\dfrac{d}{dt}
\begin{bmatrix}
 \delta \mu^1 \\
\delta \mu^2 \\
\end{bmatrix}
= -
\begin{bmatrix}
L_{11} & L_{12} \\
L_{21} & L_{22}  \\
\end{bmatrix}
\begin{bmatrix}
\delta \mu^1 \\
\delta \mu^2 \\
\end{bmatrix} .
\end{equation} 
Then, the Onsager principle posits that $L_{12}=L_{21}$. The derivation can be found in $\S \, 120$ of \citet{landau5}\footnote{Note that here we are working with the equations expressed in the ``thermodynamically conjugate form'', see equation (120.13) of \citet{landau5} and the related discussion.}.

The Onsager principle was later generalised by Casimir \cite{Onsager_Casimir}, who pointed out that the symmetry condition $L_{12}=L_{21}$ is valid only if the related dynamical variables ($\delta \mu^1$ and $\delta \mu^2$, in our case) have the same transformation properties under time reversal. Namely, if, under a transformation that conserves the positions of all the particles, but reverses the velocities \cite{Pavelka2014}, both $\delta \mu^1$ and $\delta \mu^2$ acquire the same phase ($+1$ or $-1$), then the symmetry condition $L_{12}=L_{21}$ holds. This is indeed the case for chemical potentials, which are both unchanged by a time reversal transformation. If instead one of the variables changes sign, and the other remains unchanged, then the coupling is antisymmetric: $L_{12}=-L_{21}$. Let us consider a concrete example the reader is probably familiar with.

The dynamical variables of a (macroscopic) damped harmonic oscillator immersed in an environment are the displacement $\delta x$ and the momentum $\delta p$, which both vanish at equilibrium. The entropy of a near-equilibrium state can be computed assuming that the oscillator is a point mass, with no internal degrees of freedom, so that the total entropy coincides with the entropy of the environment: $S=S_E(U_E)$. Then,  for fixed total energy $U_O+U_E$ of the isolated system ``oscillator$\, + \,$environment'', one has $T\Delta S= T\Delta S_{E}=\Delta U_E = -\Delta U_O$, where $T$ is the (fixed) temperature of the environment, and we can identify $U_O$ with the mechanical energy of the oscillator, so that
\begin{equation}
\Delta S =-\dfrac{1}{2}
\begin{bmatrix}
\delta  x &  \delta p
\end{bmatrix}
\begin{bmatrix}
kT^{-1} & 0 \\
0 & (mT)^{-1}  \\
\end{bmatrix}
\begin{bmatrix}
\delta x \\
\delta p \\
\end{bmatrix} ,
\end{equation}
where $k$ is the elastic constant and $m$ is the mass of the oscillator. We can write the equations of motion as before:
\begin{equation}\label{gorilla}
\begin{bmatrix}
kT^{-1} & 0 \\
0 & (mT)^{-1}  \\
\end{bmatrix}
\dfrac{d}{dt}
\begin{bmatrix}
\delta x \\
\delta p \\
\end{bmatrix}
= -
\begin{bmatrix}
L_{11} & L_{12} \\
L_{21} & L_{22}  \\
\end{bmatrix}
\begin{bmatrix}
\delta x \\
\delta p \\
\end{bmatrix} .
\end{equation}
Since $\delta x$ is even under time reversal, while $\delta p$ is odd, the OC principle demands that $L_{12}=-L_{21}$. Let us verify that this is reasonable. If we require that the equation of motion for $\delta x$ coincides with the definition of the canonical momentum, namely $\delta \dot{x}=\delta p/m$, then we must set $L_{11}=0$ and $L_{12}=-k(mT)^{-1}$. The OC principle then requires $L_{21}=+k(mT)^{-1}$, so that, if we introduce the notation $L_{22}=\gamma (m^2 T)^{-1}$, the second equation of \eqref{gorilla} becomes
\begin{equation}\label{hook}
\dfrac{d \delta p}{dt}=-k \delta x -\gamma \dfrac{\delta p}{m} \, ,
\end{equation}
which is indeed Newton's second law for a damped harmonic oscillator. As we can see, the Hookean force $-k \delta x$ (with the correct ``$-$'' sign) was recovered by direct application of the OC principle.

So, more in general: If the linear deviations from equilibrium of a thermodynamic system are some variables $\alpha^a$, and the entropy perturbation (truncated to second order) has the form $2\Delta S=-S_{ab}\alpha^a  \alpha^b$, where $S_{ab}=S_{ba}$ is a positive definite matrix, then, if we express the equations of motion in the form $S_{ab} \dot{\alpha}^b =-L_{ab} \alpha^b$, the OC principle demands that $L_{ab}=L_{ba} \varepsilon_a \varepsilon_b $, where $\varepsilon_a = \pm 1$ is the parity of the variable $\alpha^a$ under time reversal.

\subsection{Symmetric hyperbolicity and well-posedness}

Let us now turn our attention to the concept of ``symmetric hyperbolicity''. Again, the best place to start is by considering a concrete example. Let $\delta\rho(t,x)$ and $\delta u(t,x)$ be the linear perturbations to the energy density and the flow velocity of a fluid near homogeneous thermodynamic equilibrium (in 1+1 dimensions, for simplicity). Working in some natural units, consider the two alternative oversimplified models below:
\begin{equation}\label{Model12}
\begin{split}
& \text{Model 1:} \quad 
\begin{bmatrix}
\partial_t & \partial_x \\
\partial_x & \partial_t \\
\end{bmatrix}
\begin{bmatrix}
\delta \rho \\
\delta u \\
\end{bmatrix}=0 \, , \\
& \text{Model 2:} \quad
\begin{bmatrix}
\partial_t & \partial_x \\
-\partial_x & \partial_t \\
\end{bmatrix}
\begin{bmatrix}
\delta \rho \\
\delta u \\
\end{bmatrix} =0 \, .\\
\end{split}
\end{equation} 
Which model is more ``reasonable''? To answer this question, one just needs to contract (on the left) both sides of both models with the row matrix 
$\begin{bmatrix}
\partial_t & -\partial_x
\end{bmatrix}$. The result is the following:
\begin{equation}\label{models12}
\begin{split}
& \text{Model 1:} \quad (\partial^2_t -\partial^2_x)\delta \rho=0 \, ,\\
& \text{Model 2:} \quad (\partial^2_t +\partial^2_x)\delta \rho =0 \, .\\
\end{split}
\end{equation}
According to Model 1, sound propagates in accordance with the wave equation, while Model 2 is more unconventional, positing that sound waves are governed by the Laplace equation. Now, we know from physical experience that Model 1 is the correct one. We can also argue that Model 2 is not acceptable on a thermodynamic basis, since thermodynamic equilibrium is stable by definition \cite{Hiscock_Insatibility_first_order,GavassinoBounds2023}, while the Laplace equation admits growing Fourier modes with dispersion relation $\omega=i|k|$. However, Model 2 is also problematic from a purely mathematical perspective, as it constitutes the prototype of an illposed initial value problem. To understand this, consider the following reasoning (due to Hadamard \cite{Joseph1990}). The second equation of \eqref{models12} admits solutions of the form 
\begin{equation}
\delta \rho_n(t,x)= e^{-\sqrt{n}}\cos(nx)\cosh(nt) \, ,
\end{equation} 
for every positive integer $n$. Let us then take the limit as $n \rightarrow +\infty$. At $t=0$, these solutions, and all their derivatives, converge uniformly to zero, since $\partial^\alpha\delta \rho_n(0,x) \sim n^\alpha e^{-\sqrt{n}} \rightarrow 0$. On the other hand, for any $t>0$, these solutions tend to oscillate with infinitely large amplitude, since $\delta \rho_n(t,x) \sim e^{nt-\sqrt{n}}\rightarrow \infty$. Hence, the instability is infinitely fast (there is no universal Lyapunov exponent controlling the growth), and one can construct arbitrarily large solutions from arbitrarily small initial data. In this case, we say that the solution does not ``depend continuously on the data''. Another problem of Model 2 is the fact that we are not free to impose arbitrary (smooth) initial conditions for $\delta \rho$ and $\delta u$ at $t=0$. In fact, if we define the complex function $f=\delta \rho + i \delta u$, we find that it satisfies the Cauchy Riemann equation $(\partial_t -i\partial_x)f=0$ \cite{Benzoni_book}, and it can be shown (see \S 1.1 of \citet{Rauch_book}) that solutions exist if and only if $f(0,x)$ is analytic. This also implies that Model 2 is strongly non-local (and therefore acausal), because knowledge of the initial data $f(0,x)$ in a finite interval of space uniquely fixes the function $f(0,x)$ everywhere.

The mathematical theory of partial differential equations calls Model 1 ``symmetric hyperbolic'' and Model 2 ``elliptic''. While the latter is plagued (as we have just shown) by a number of non-desirable features, it can be shown that none of these issues can occur in the former. In fact, all symmetric hyperbolic systems are strongly hyperbolic, and this implies that they have wellposed initial value problems in the linear regime, i.e. the solution exists for arbitrary smooth initial data, it is unique, and it depends continuously on the given data \cite{Kato2009}. The general definition of a symmetric hyperbolicity is the following. Given a collection of fields $\varphi^A(x^\mu)$, a (linear, in our case) system of partial differential equations of the form $M^\mu_{AB}\partial_\mu \varphi^B =-\Xi_{AB}\varphi^B$ is said to be symmetric hyperbolic if
\begin{itemize}
\item All matrices $M^\mu_{AB}$ are symmetric in $A$ and $B$;
\item The matrix $M^0_{AB}$ is positive definite.
\end{itemize}
This is indeed the case of Model 1, for which $\varphi^A=\{\delta \rho, \delta u \}$,  and 
\begin{equation}
M^0_{AB} =
\begin{bmatrix}
1 & 0 \\
0 & 1 \\
\end{bmatrix} ,
\spc 
M^1_{AB} =
\begin{bmatrix}
0 & 1 \\
1 & 0 \\
\end{bmatrix} .
\end{equation}  
Note that the assumption that $M^0_{AB}$ is positive definite cannot be released. In fact, take Model 2, and multiply (on the left) both sides of the system \eqref{Model12} by the Pauli matrix $\sigma_3$. The result is
\begin{equation}
\begin{bmatrix}
\partial_t & \partial_x \\
\partial_x & -\partial_t \\
\end{bmatrix}
\begin{bmatrix}
\delta \rho \\
\delta u \\
\end{bmatrix} =0 \, .\\
\end{equation}
This is indeed a symmetric system, but it is not symmetric hyperbolic (actually, it is still elliptic), because the matrix $M^0_{AB}=\text{diag}(1,-1)$ is not positive definite. 

The bottom line of this section is that the sign of a coefficient affects not only the physical behavior, but also the mathematical properties of a fluid theory. If we turn a plus into a minus, the resulting equations may become illposed.

\subsection{Symmetric hyperbolicity from the Onsager-Casimir principle}

The question that we want to answer is the following: Can we ``choose'' Model 1 over Model 2 in equation \eqref{Model12} using \textit{only} the OC principle, and without assuming any other knowledge? More precisely, let us consider an agnostic model of the form
\begin{equation}\label{agnostuz}
\begin{bmatrix}
\partial_t & \partial_x \\
\sigma\partial_x & \partial_t \\
\end{bmatrix}
\begin{bmatrix}
\delta \rho \\
\delta u \\
\end{bmatrix} =0 \, ,\\
\end{equation}
where $\sigma$ is an undetermined constant. Can we fix the value of $\sigma$ in the same way as the Hookean force $-k\delta x$ in \eqref{hook} was fixed using the OC principle? The main difficulty that we encounter is that the OC principle deals with ordinary differential equations involving only time, while here we are dealing with partial differential equations involving both time and space. Luckily, this problem is easily solved if we decompose the fields into Fourier modes. In fact, each individual Fourier mode decouples from all other Fourier modes, and it contributes to the dynamics with a separate set of degrees of freedom, which are governed by an independent system of ordinary differential equations\footnote{The idea of applying the OC principle to a field theory in the Fourier space is not new \cite{Geigenmuller1983}. \citet{Onsager_Casimir} himself applied the OC principle to the conduction of heat in crystals, treating the Fourier components of the temperature as the fundamental degrees of freedom.}. In particular, if we focus on a single mode, with wavenumber $k=1$, we have that
\begin{equation}\label{dozensss}
\begin{split}
& \delta \rho(t,x) = \rho_s(t) \sin(x)+ \rho_c(t) \cos(x) \, ,\\
& \delta u(t,x) = u_s(t)\sin(x) +u_c(t) \cos(x) \, , \\
\end{split}
\end{equation}
and we can treat $\{\rho_s,u_s,\rho_c,u_c\}$ as our dynamical variables $\alpha^a$. Plugging \eqref{dozensss} into \eqref{agnostuz}, we obtain a closed system:
\begin{equation}\label{quebueq}
\dfrac{d}{dt}
\begin{bmatrix}
\rho_s \\
u_s \\
\rho_c \\
u_c \\
\end{bmatrix} =- 
\begin{bmatrix}
0 & 0 & 0 & -1 \\
0 & 0 & -\sigma & 0 \\
0 & 1 & 0 & 0 \\
\sigma & 0 & 0 & 0 \\
\end{bmatrix} 
\begin{bmatrix}
\rho_s \\
u_s \\
\rho_c \\
u_c \\
\end{bmatrix} . 
\end{equation}
Concerning the entropy perturbation, we can reasonably assume that it is given by a volume integral of the form $2\Delta S=-\int (\delta \rho^2 +\delta u^2)dx$. Again, to keep the discussion simple, we have assumed that we could set all dimensionality constants to one by appropriately fixing the units (this is just a toy-model). Plugging \eqref{dozensss} into the integral formula for the entropy perturbation, we find that $\Delta S \propto \rho_s^2+u_s^2+\rho_c^2+u_c^2$. Note that the double products $2\rho_s \rho_c$ and $2u_s u_c$ do not appear because the function $\sin(x)\cos(x)$ has zero average when integrated over the period. Thus, the matrix $S_{ab}$ is proportional to the $4 \times 4$ identity matrix, meaning that equation \eqref{quebueq} is already written in the form $S_{ab} \dot{\alpha}^b =-L_{ab} \alpha^b$, up to an irrelevant multiplicative constant\footnote{Indeed, note that, if we replace the ``canonical'' form $S_{ab} \dot{\alpha}^b =-L_{ab} \alpha^b$ with a rescaled version of it, $zS_{ab} \dot{\alpha}^b =-zL_{ab} \alpha^b$ (with $z\neq 0$), the symmetry principle also applies to the matrix $zL_{ab}$. In fact, multiplying the equation $L_{ab}=L_{ba} \varepsilon_a \varepsilon_b $ by $z$, we get $zL_{ab}=zL_{ba} \varepsilon_a \varepsilon_b$.}. Therefore, we can just impose the OC symmetry conditions on the $4 \times 4$ matrix on the right-hand side of \eqref{quebueq}. We immediately note that such matrix couples only ``$\rho$'' terms with ``$u$'' terms (and vice versa), which have opposite parity under time reversal\footnote{The variables $\varphi^A_s(t)$ and $\varphi^A_c(t)$ acquire under time reversal the same phase $\varepsilon_A$ as the field $\varphi^A(t,x)$. To prove this, one can just notice that $\varphi^A_s=\varphi^A(x=\pi/2)$ and $\varphi^A_c=\varphi^A(x=0)$.}, the energy being even ($\varepsilon_\rho=+1$) and velocity being odd ($\varepsilon_u=-1$). Thus, we are in a situation that is analogous that of the damped harmonic oscillator, and we need to impose an antisymmetric coupling: $\sigma {=}L_{u_c \rho_s}\overset{\text{oc}}{=} L_{\rho_s u_c}\varepsilon_\rho \varepsilon_u{=}-L_{\rho_s u_c}{=}-(-1)$. This leads us to the condition $\sigma=+1$, i.e. Model 1. As we can see, symmetric hyperbolicity follows directly from the OC principle.

\section{Israel-Stewart theory in 1+1 dimensions}\label{IS11}

The surprising agreement between the OC principle and symmetric hyperbolicity found in the example above seems to be a happy coincidence. After all, equation \eqref{agnostuz} is an oversimplified toy model. Let us consider a ``more complicated'' theory, and see if we find similar results. Here, we will focus on the linearised Israel-Stewart theory in the Eckart frame \cite{Israel_Stewart_1979,Hishcock1983}, in 1+1 dimensions, and in the absence of shear viscosity. We include, as dissipative processes, bulk viscosity and heat conduction.

\subsection{Symmetric hyperbolic form of the equations}

The linearised degrees of freedom of the theory can be taken to be the fields $\varphi^A=\{T\delta \alpha, \beta\delta T, \delta \Pi,\delta u, \delta q \}$, representing perturbations to respectively the fugacity (rescaled by the background temperature $T$), the temperature (rescaled by the background inverse temperature, $\beta=T^{-1}$), the bulk viscous stress, the flow velocity, and the heat flux. \citet{Hishcock1983} have shown that, with this choice of variables, the field equations $M^\mu_{AB}\partial_\mu \varphi^B=-\Xi_{AB}\varphi^B$ can be written in a manifestly symmetric form as follows:
\begin{equation}\label{astolfo}
\begin{bmatrix}
\beta n_{,\alpha} \partial_t & \beta \rho_{,\alpha} \partial_t & 0 & n\partial_x & 0  \\
 \beta \rho_{,\alpha} \partial_t & T \rho_{,T} \partial_t & 0 & w\partial_x & \partial_x  \\
0 & 0 & b_0 \partial_t & \partial_x & -a_0 \partial_x \\
n \partial_x & w\partial_x & \partial_x & w\partial_t & \partial_t \\
0 & \partial_x & -a_0 \partial_x & \partial_t & b_1 \partial_t \\
\end{bmatrix}
\begin{bmatrix}
T\delta \alpha \\
\beta \delta T \\
\delta \Pi \\
\delta u \\
\delta q \\
\end{bmatrix} =- 
\begin{bmatrix}
0 \\
0 \\
\delta \Pi/\zeta \\
0 \\
\beta\delta q/\kappa \\
\end{bmatrix},
\end{equation} 
where $n$, $w$, $\kappa$, $\zeta$, $b_0$, $b_1$, and $\alpha_0$ are respectively the (background) particle density, enthalpy density, heat conductivity, bulk viscosity, and ``second-order'' transport coefficients \cite{DMNR2012}. The symbols $n_{,\alpha}$, $\rho_\alpha$, and $\rho_{,T}$ denote partial derivatives of the thermodynamic functions $n(\alpha,T)$ and $\rho(\alpha,T)$. As it turns out, the system \eqref{astolfo} is not only symmetric, but also symmetric hyperbolic, if the theory is thermodynamically stable \cite{GavssinoUniversalityI2023,GavssinoUniversalityII2023}. In fact, the information current $\mathbb{E}^\mu$ of the Israel-Stewart theory \cite{GavassinoGibbs2021} can be expressed (in 1+1 dimensions) as follows:
\begin{equation}\label{astolfo2}
\begin{split}
T\mathbb{E}^0 ={}&\dfrac{1}{2}\bigg[ \beta n_{,\alpha}(T\delta \alpha)^2 + 2\beta \rho_{,\alpha}(T\delta \alpha)(\beta \delta T)+T\rho_{,T}(\beta \delta T)^2+b_0 (\delta \Pi)^2 +w (\delta u)^2+2\delta u \delta q +b_1 (\delta q)^2 \bigg] \, , \\
T\mathbb{E}^1 ={}& (w \beta \delta T +nT\delta \alpha+\delta \Pi)\delta u + (\beta \delta T - a_0 \delta \Pi)\delta q \, .  \\
\end{split}
\end{equation}
Comparing \eqref{astolfo} with \eqref{astolfo2}, it is immediate to see that $2T\mathbb{E}^\mu=M^\mu_{AB}\varphi^A \varphi^B$. Hence, if the information current is future directed timelike for any $\varphi^A \neq 0$ (which is the condition for thermodynamic stability \cite{GavassinoCausality2021}), it automatically follows that $M^0_{AB}$ is positive definite, and thus the theory is symmetric hyperbolic.

It is worth stressing that the symmetric structure of \eqref{astolfo} is of interest not only for its mathematical implications, but it is also a genuine physical prediction of the theory. In fact, consider the transport coefficient $a_0$, which describes the coupling between bulk viscosity and heat conduction. The fact that it is the same coefficient appearing both in the third and in the fifth line of \eqref{astolfo} implies that, if the presence of $\delta q$ ``affects'' the relaxation of $\delta \Pi$, then exactly in the same way the presence of $\delta \Pi$ must ``affect'' the relaxation of $\delta q$, with the same coupling coefficient $a_0$. Indeed, \citet{Liu1986} (section 7.6) have explicitly verified that this symmetry emerges in quite a miraculous way also in kinetic theory. This should not be surprising, since it is the kind reciprocal relation that we expect to follow from the OC principle. Let us then pursue this intuition. 

\subsection{The Onsager-Casimir principle in the Israel-Stewart theory}

As we did before, we focus on a single Fourier mode with $k=1$, and we decompose the fields as follows:
\begin{equation}\label{gzimobo}
\begin{split}
 \delta \alpha(t,x)={}& \alpha_s(t)\sin(x)+\alpha_c(t) \cos(x) \, ,\\
 \delta T(t,x)={}&  T_s(t)\sin(x)+ T_c(t) \cos(x) \, ,\\
 \delta \Pi(t,x)={}&  \Pi_s(t)\sin(x)+ \Pi_c(t) \cos(x) \, ,\\
\delta u(t,x)={}&  u_s(t)\sin(x)+ u_c(t) \cos(x) \, ,\\
 \delta q(t,x)={}&  q_s(t)\sin(x)+ q_c(t) \cos(x) \, ,\\
\end{split}
\end{equation}
and we take, as dynamical variables, $\alpha^a=\{T\alpha_s,\beta T_s,\Pi_s,u_s,q_s,T\alpha_c,\beta T_c, \Pi_c,u_c,q_c \}$. The entropy perturbation is given by $\Delta S=-\int \mathbb{E}^0 dx$, see \cite{GavassinoGibbs2021} for the proof. This allows us to compute the matrix $S_{ab}$, which is given by
\begin{equation}
S_{ab} \propto 
\left[\begin{array}{@{}ccccc|ccccc@{}}
\beta n_{,\alpha} & \beta \rho_{,\alpha} & 0 & 0 & 0 & 0 & 0 & 0 & 0 & 0\\
\beta \rho_{,\alpha} & T\rho_{,T} & 0 & 0 & 0 & 0 & 0 & 0 & 0 & 0 \\
0 & 0 & b_0 & 0 & 0 & 0 & 0 & 0 & 0 & 0 \\
0 & 0 & 0  & w & 1 & 0 & 0 & 0 & 0 & 0 \\
0 & 0 & 0 & 1 & b_1 & 0 & 0 & 0 & 0 & 0 \\ \hline
0&0&0&0&0&\beta n_{,\alpha} & \beta \rho_{,\alpha} & 0 & 0 & 0 \\
0&0&0&0&0&\beta \rho_{,\alpha} & T\rho_{,T} & 0 & 0 & 0 \\
0&0&0&0&0&0 & 0 & b_0 & 0 & 0 \\
0&0&0&0&0&0 & 0 & 0  & w & 1 \\
0&0&0&0&0&0 & 0 & 0 & 1 & b_1 \\
\end{array}\right] ,
\end{equation}
where the proportionality constant scales like the volume. We have divided the above matrix into four blocks to separate the ``sine variables'' from the ``cosine variables''. Plugging \eqref{gzimobo} into \eqref{astolfo}, we can express the equations of motion in the canonical form $S_{ab}\dot{\alpha}^b=-L_{ab}\alpha^b$, with
\begin{equation}\label{sangria}
L_{ab} \propto 
\left[\begin{array}{@{}ccccc|ccccc@{}}
\textcolor{blue}{0}& \textcolor{blue}{0}& \textcolor{blue}{0}& \textcolor{red}{0}& \textcolor{red}{0}& \textcolor{blue}{0} & \textcolor{blue}{0} & \textcolor{blue}{0} & \textcolor{red}{-n} & \textcolor{red}{0} \\
\textcolor{blue}{0}& \textcolor{blue}{0}& \textcolor{blue}{0}& \textcolor{red}{0}& \textcolor{red}{0}& \textcolor{blue}{0} & \textcolor{blue}{0} & \textcolor{blue}{0} & \textcolor{red}{-w} & \textcolor{red}{-1} \\
\textcolor{blue}{0}& \textcolor{blue}{0}& \textcolor{blue}{\zeta^{-1}}& \textcolor{red}{0}& \textcolor{red}{0}& \textcolor{blue}{0} & \textcolor{blue}{0} & \textcolor{blue}{0} & \textcolor{red}{-1} & \textcolor{red}{a_0} \\
\textcolor{red}{0}& \textcolor{red}{0}& \textcolor{red}{0}& \textcolor{blue}{0}& \textcolor{blue}{0}& \textcolor{red}{-n} & \textcolor{red}{-w} & \textcolor{red}{-1} & \textcolor{blue}{0} & \textcolor{blue}{0} \\
\textcolor{red}{0}& \textcolor{red}{0}& \textcolor{red}{0}& \textcolor{blue}{0}& \textcolor{blue}{\beta/\kappa}& \textcolor{red}{0} & \textcolor{red}{-1} & \textcolor{red}{a_0} & \textcolor{blue}{0} & \textcolor{blue}{0} \\ \hline
\textcolor{blue}{0}& \textcolor{blue}{0}& \textcolor{blue}{0}& \textcolor{red}{n}& \textcolor{red}{0}& \textcolor{blue}{0} & \textcolor{blue}{0} & \textcolor{blue}{0} & \textcolor{red}{0} & \textcolor{red}{0} \\
\textcolor{blue}{0}& \textcolor{blue}{0}& \textcolor{blue}{0}& \textcolor{red}{w}& \textcolor{red}{1}& \textcolor{blue}{0} & \textcolor{blue}{0} & \textcolor{blue}{0} & \textcolor{red}{0} & \textcolor{red}{0} \\
\textcolor{blue}{0}& \textcolor{blue}{0}& \textcolor{blue}{0}& \textcolor{red}{1}& \textcolor{red}{-a_0}& \textcolor{blue}{0} & \textcolor{blue}{0} & \textcolor{blue}{\zeta^{-1}} & \textcolor{red}{0} & \textcolor{red}{0} \\
\textcolor{red}{n}& \textcolor{red}{w}& \textcolor{red}{1}& \textcolor{blue}{0}& \textcolor{blue}{0}& \textcolor{red}{0} & \textcolor{red}{0} & \textcolor{red}{0} & \textcolor{blue}{0} & \textcolor{blue}{0} \\
\textcolor{red}{0}& \textcolor{red}{1}& \textcolor{red}{-a_0}& \textcolor{blue}{0}& \textcolor{blue}{0}& \textcolor{red}{0} & \textcolor{red}{0} & \textcolor{red}{0} & \textcolor{blue}{0} & \textcolor{blue}{\beta/\kappa} \\
\end{array}\right] .
\end{equation}
In the above matrix, we have marked in blue the elements that should be symmetric according to the OC principle, and in red those that should be antisymmetric. As can be seen, the Israel-Stewart theory is fully consistent with the OC principle. In particular, its symmetric hyperbolic character is a manifestation of the Onsager reciprocal relations, which are so restrictive that they constrain the very mathematical structure of the theory. The symmetry of the coupling $a_0$ is also a consequence of the OC principle. In fact, as we can see from equation \eqref{sangria}, the dynamical equation for $\Pi_s$ (third line) involves the variable $q_c$ (tenth column), and at the same time the dynamical equation for $q_c$ (tenth line) involves the variable $\Pi_s$ (third column). But since $\{\Pi_s,q_c \}$ is an even-odd couple, the corresponding coupling must be antisymmetric, enforcing the corresponding coefficients $a_0$ in \eqref{astolfo} to be the same.

\section{General argument in 3+1 dimensions}

Our goal, now, is to generalise the result of the previous section to an arbitrary hydrodynamic system in 3+1 dimensions. We will show that, in most circumstances, the OC principle implies that linear perturbations about homogeneous equilibria are governed by a symmetric-hyperbolic system of equations. The reasoning is essentially the same as above, with some little caveats.

\subsection{Conjugate variables}

Consider an isolated hydrodynamic system, and assume that its macroscopic state can be characterised by a finite collection of macroscopic (real) tensor fields $\varphi^A(t,\textbf{x})$, where $A$ is a multi-index. For convenience, we construct these fields in such a way that $\varphi^A=0$ at equilibrium, and we work in the equilibrium global rest frame (we consider only homogeneous equilibria). Then, the total entropy of a state differing from the equilibrium state is $S=S_\text{eq}+\Delta S$, with
\begin{equation}\label{DeltaS}
\Delta S = -\dfrac{1}{2} \int \mathcal{S}_{AB} \,  \varphi^A  \varphi^B d^3 x +\mathcal{O}( \varphi\varphi\varphi) + \mathcal{O}(\partial ) \, ,
\end{equation}
where $\mathcal{S}_{AB}$ is a positive-definite (constant) symmetric matrix. We work in the limit of small perturbations ($\varphi^A \rightarrow 0$) and small wavenumbers ($\partial \rightarrow 0$), so that we can neglect all the corrections that scale like $\mathcal{O}( \varphi\varphi\varphi)$ and $\mathcal{O}(\partial )$. Like before, let us focus on the dynamics of a single Fourier mode, with wavenumber $\textbf{k}=(k_x,k_y,k_z)$, in the limit as $\textbf{k}\rightarrow 0$. Then, the field components $\varphi^A$ can be decomposed into $2^3=8$ pieces:
\begin{equation}\label{Deccoppola}
\begin{split}
\varphi^A(t,\textbf{x})&{}= \varphi^A_1(t) \sin(k_x x)\sin(k_y y)\sin(k_z z) + \varphi^A_2(t) \cos(k_x x)\sin(k_y y)\sin(k_z z) \\
& +\varphi^A_3(t) \sin(k_x x)\cos(k_y y)\sin(k_z z) +\varphi^A_4(t) \sin(k_x x)\sin(k_y y)\cos(k_z z) \\
& +\varphi^A_5(t) \cos(k_x x)\cos(k_y y)\sin(k_z z) +\varphi^A_6(t) \cos(k_x x)\sin(k_y y)\cos(k_z z) \\
& +\varphi^A_7(t) \sin(k_x x)\cos(k_y y)\cos(k_z z) +\varphi^A_8(t) \cos(k_x x)\cos(k_y y)\cos(k_z z) \, .\\
\end{split}
\end{equation} 
If we plug this field configuration into \eqref{DeltaS}, we get
\begin{equation}\label{Grinuzzo}
\Delta S \propto -\dfrac{1}{2} \sum_{n=1}^8 \mathcal{S}_{AB} \,  \varphi^A_n  \varphi^B_n  \, ,
\end{equation}
because the Fourier ``sine-cosine basis'' is orthonormal with respect to volume-integration. Now, let us define the ``conjugate fields'' $\mathbb{A}_A:= \mathcal{S}_{AB}\varphi^B$. If we perform for $\mathbb{A}_A(t,\textbf{x})$ the same decomposition that we made for $\varphi^A(t,\textbf{x})$ in equation \eqref{Deccoppola}, we obtain a collection of 8 time-dependent variables:
\begin{equation}
\mathbb{A}^n_A(t) = \mathcal{S}_{AB} \, \varphi^B_n(t) \spc (n=1,...,8) \, .
\end{equation} 
Because of equation \eqref{Grinuzzo}, these variables are the thermodynamic conjugates of $\varphi^A_n$, in the
sense that \cite{landau5} 
\begin{equation}\label{ConjugoA}
\braket{\mathbb{A}_A^n(t) \, \varphi^B_m(t)}= \text{const} \times\delta\indices{^B _A} \delta\indices{^n _m}  \, ,
\end{equation}
where $\braket{...}$ denotes the microcanonical average (which is the appropriate ensemble average for an isolated system \cite{huang_book}).

\subsection{From field equations to ordinary differential equations}

Since the fields $\varphi^A(t,\textbf{x})$ characterise the macrostate of the system completely at a time $t$, and since the matrix $\mathcal{S}_{AB}$ is invertible, the equations of motion of the system can be expressed in the form
\begin{equation}
\partial_t \mathbb{A}_A(t,\textbf{x})=-F_A[\varphi^B(t,\textbf{y});\textbf{x}] \, ,
\end{equation} 
where $F_A[\varphi^B(t,\textbf{y});\textbf{x}]$ are some linear functionals of $\varphi^B(t,\textbf{y})$ at that given time $t$. By locality, these functionals should depend only on the value the fields and of their spatial derivatives (of arbitrary order) at $\textbf{x}$ \cite{GavassinoFronntiers2021}, namely $F_A=F_A(\varphi^B, \partial_j \varphi^B,\partial_j \partial_k \varphi^B,...)$. Given that we are working in the limit of small spatial gradients, we can gradient-expand $F_A$, and truncate the result to first order:
\begin{equation}\label{LeFields}
\partial_t \mathbb{A}_A = -\Xi_{AB} \, \varphi^B -M^j_{AB} \, \partial_j \varphi^B + \mathcal{O}(\partial^2) \, .
\end{equation}
These can be interpreted as the field equations of the linearised hydrodynamic theory.
We are not allowed to truncate the field equations at higher orders in the spatial gradients because they would be acausal \cite{CourantHilbert2_book} and, therefore, non-exploitable in a relativistic context \cite{GavassinoCausality2021,
GavassinoSuperluminal2021,GavassinoStabilityCarter2022}. Now, since we are interested in the dynamics of a single Fourier mode, we can just plug \eqref{Deccoppola} into \eqref{LeFields}, and this produces the system of ordinary differential equations below:
\begin{equation}\label{ordinariocomeilgrasso}
\dfrac{d}{dt} 
\begin{bmatrix}
\mathbb{A}_A^1 \\[3pt]
\mathbb{A}_A^2 \\[3pt]
\mathbb{A}_A^3 \\[3pt]
\mathbb{A}_A^4 \\[3pt]
\mathbb{A}_A^5 \\[3pt]
\mathbb{A}_A^6 \\[3pt]
\mathbb{A}_A^7 \\[3pt]
\mathbb{A}_A^8 \\[3pt]
\end{bmatrix}
= -
\begin{bmatrix}
\Xi_{AB} & -k_x M^x_{AB} & -k_y M^y_{AB} & -k_z M^z_{AB} & 0 & 0 & 0 & 0\\[3pt]
k_x M^x_{AB} & \Xi_{AB}  & 0 & 0 & -k_y M^y_{AB} & -k_z M^z_{AB} & 0 & 0\\[3pt]
k_y M^y_{AB} & 0 & \Xi_{AB}  & 0 & -k_x M^x_{AB} & 0 & -k_z M^z_{AB} & 0\\[3pt]
k_z M^z_{AB} & 0 & 0 & \Xi_{AB}  & 0 & -k_x M^x_{AB} & -k_y M^y_{AB} & 0\\[3pt]
0 & k_y M^y_{AB} & k_xM^x_{AB} & 0 & \Xi_{AB}  & 0 & 0 & -k_z M^z_{AB} \\[3pt]
0 & k_z M^z_{AB} & 0 & k_x M^x_{AB} & 0 & \Xi_{AB}  & 0 & -k_y M^y_{AB}\\[3pt]
0 & 0 & k_z M^z_{AB} & k_y M^y_{AB} & 0 & 0 & \Xi_{AB}  & -k_x M^x_{AB}\\[3pt]
0 & 0 & 0 & 0 & k_z M^z_{AB} & k_y M^y_{AB} & k_x M^x_{AB} & \Xi_{AB} \\[3pt]
\end{bmatrix}
\begin{bmatrix}
\varphi_1^B \\[3pt]
\varphi_2^B \\[3pt]
\varphi_3^B \\[3pt]
\varphi_4^B \\[3pt]
\varphi_5^B \\[3pt]
\varphi_6^B \\[3pt]
\varphi_7^B \\[3pt]
\varphi_8^B \\[3pt]
\end{bmatrix} \, .
\end{equation}
This system generalises equation \eqref{quebueq} to an arbitrary fluid in 3+1 dimensions. Note that, if $\mathfrak{D}$ is the number of independent components of the fields $\varphi^A$, then this is a system of $8\mathfrak{D}$ ordinary differential equations, and the square matrix on the right-hand side can be viewed as an $8\mathfrak{D}\times 8\mathfrak{D}$ matrix, with $8\times 8$ blocks having size $\mathfrak{D}\times \mathfrak{D}$ each.

\subsection{Implications of the Onsager-Casimir principle}

Recalling equation \eqref{ConjugoA}, we see that the system \eqref{ordinariocomeilgrasso} is written in Casimir's form \cite{landau5}: the time-derivatives of some non-equilibrium variables are expressed in terms of their conjugate variables. Then, we can apply the OC principle to the $8\mathfrak{D}\times 8\mathfrak{D}$ matrix in equation \eqref{ordinariocomeilgrasso}, and we obtain the conditions
\begin{equation}\label{gabumbo}
 \Xi_{BA}= \Xi_{AB} \, \varepsilon_A \varepsilon_B \, , \spc
 M^j_{BA}= {-}M^j_{AB} \, \varepsilon_A \varepsilon_B \, .
\end{equation}
where $\varepsilon_A$ is the phase that the field component $\varphi^A$ acquires under time reversal (which coincides with the phase acquired by $\mathbb{A}_A$ \cite{Onsager_Casimir}). The implications of the first equation of \eqref{gabumbo} have been extensively discussed in \cite{GavassinoNonHydro2022}. The second equation is what we are interested in, now. But before discussing its implications, let us convince ourselves that it is correct. 
There are three ways of deriving equation \eqref{gabumbo}. The most rigorous way is to express the OC principle using matrix notation, and apply it to the $8\mathfrak{D}\times 8\mathfrak{D}$ block matrix in \eqref{ordinariocomeilgrasso}. We do this in Appendix \ref{camusso}. The second way (which is less rigorous, and requires a bit of faith) is to extend the OC symmetry relations to field theories following a reasoning that is analogous to that of \citet{Onsager_Casimir}. We do this in Appendix \ref{AAA}. The third (and most direct) way is to verify that equation \eqref{gabumbo} is true with an explicit example. Here, we will follow the third way. 

Let us say, for clarity, that the number of independent field components is  $\mathfrak{D}=14$, so that the multi-index $A$ runs from 1 to 14. Then, we can write out the evolution equations for the variables, say, $\mathbb{A}^2_9$ and $\mathbb{A}^5_{13}$ [see equation \eqref{ordinariocomeilgrasso}]:
\begin{equation}
\begin{split}
\dot{\mathbb{A}}^2_9 ={}& -k_x M^x_{9,B} \, \varphi^B_1 - \Xi_{9,B} \, \varphi_2^B + k_y M^y_{9,B} \, \varphi^B_5 + k_z M^z_{9,B} \, \varphi^B_6 \, , \\
\dot{\mathbb{A}}^5_{13} ={}& -k_y M^y_{13,B} \, \varphi_2^B -k_x M^x_{13,B}\, \varphi_3^B - \Xi_{13,B} \, \varphi^B_5 + k_z M^z_{13,B}\, \varphi_8^B \, .  \\
\end{split}
\end{equation} 
To apply the OC principle, we only need to find in the two equations above the coupling between  $\mathbb{A}^2_9$ and $\mathbb{A}^5_{13}$. That is, we need to find the term proportional to $\varphi_5^{13}$ (the conjugate to $\mathbb{A}^5_{13}$) in the evolution equation of $\mathbb{A}^2_9$, and vice versa. This is easily done:
\begin{equation}
\begin{split}
\dot{\mathbb{A}}^2_9 ={}& +k_y M^y_{9,13} \, \varphi^{13}_5 +\text{``All the rest''} \, , \\
\dot{\mathbb{A}}^5_{13} ={}& -k_y M^y_{13,9} \, \varphi_2^9 +\text{``All the rest''} \, .  \\
\end{split}
\end{equation} 
Then, the OC principle tells us that $k_y M^y_{9,13}=(-k_y M^y_{13,9})\varepsilon_9 \varepsilon_{13}$. Simplifying $k_y$, we recover the second equation of \eqref{gabumbo}, for $j=y$, $A=9$, and $B=13$. Repeating this procedure for all possible couples $\{\mathbb{A}_A^n, \mathbb{A}_B^m \}$, one obtains \eqref{gabumbo}. 

Now we are ready to study the consequences of the second equation of \eqref{gabumbo}. The symmetry properties of $M^j_{AB}$ are uniquely determined by the sign of the product $\varepsilon_A\varepsilon_B$. In a hydrodynamic theory, there are quantities that are even under time reversal (e.g. the energy density) and quantities that are odd (e.g the flow velocity). Therefore, the product $\varepsilon_A \varepsilon_B$ can be both positive and negative, depending on $A$ and $B$. However, we can make a useful observation. Normally, fluids are isotropic at equilibrium. Therefore, the matrix $M^j_{AB}$ cannot identify a preferred direction in space, and it must be an isotropic tensor, with respect to rotations of space. In the table below, we show the most general form that $M^j_{AB}$ can have, depending on the transformation properties of $\varphi^A$ and $\varphi^B$ under 3D-rotations \cite{Kearsley1975}.
\renewcommand{\arraystretch}{1.5}
\begin{center}
\begin{tabular}{ |l||l|l|l| } 
 \hline
`` Table of $M^j_{AB}$ ''  & Scalar: $\varphi^B=\psi$ & Vector: $\varphi^B=\psi^h$ & 2-Tensor: $\varphi^B=\psi^{hr}$ \\ \hline \hline
Scalar: $\varphi^A=\varphi$ & 0 & $\alpha \, \delta\indices{^j _h}$ & $\alpha \, \epsilon\indices{^j _h _r}$  \\ \hline
Vector: $\varphi^A=\varphi^k$ & $\alpha \, \delta\indices{^j _k}$ & $\alpha \, \epsilon\indices{^j _k _h}$ &  $\alpha \, \delta\indices{^j _k}\delta_{hr} + \beta \, \delta\indices{^j _h}\delta_{kr}+\gamma \, \delta\indices{^j _r}\delta_{hk}$  \\ \hline
2-Tensor: $\varphi^A=\varphi^{kl}$ &  $\alpha \, \epsilon\indices{^j _k _l}$ & $\alpha \, \delta\indices{^j _k}\delta_{lh} + \beta \, \delta\indices{^j _h}\delta_{lk}+\gamma \, \delta\indices{^j _l}\delta_{hk}$ & $\alpha \, \epsilon\indices{^j _k _l} \delta_{hr} + \text{``similar terms''}$  \\ 
 \hline
\end{tabular}
\end{center}
Now, the only hydrodynamic systems that present the Levi-Civita tensor $\epsilon\indices{^j _k _h}$ in the field equations are magnetised fluids, or fluids with spin. It is well-known that, in these cases, one needs to have extra care with applying the OC principle. For this reason, we will ignore this possibility here, and we will restrict our attention to non-magnetised spinless fluids, so that all the terms with the Levi-Civita tensor in the table are set to zero. But, then, $M^j_{AB} \neq 0$ only if it couples a field with an even number of spacelike indices (e.g. $\varphi$, or $\varphi^{kl}$) to a field with an odd number of spacelike indices (e.g. $\psi^h$). On the other hand, fields with an even number of spacelike indices (e.g. temperature, densities, and stresses) are usually even under time reversal, while fields with an odd number of spacelike indices (e.g. velocity, heat flux, and diffusive currents) are usually odd under time reversal\footnote{The only exceptions the author is aware of are the ``holographic partners'' introduced in \cite{GavassinoNonHydro2022}. In theories that contain such partners, equation \eqref{SymmetryForever} may be in principle violated. However, it should be noted that, in the theories constructed by \citet{GavassinoNonHydro2022}, all the matrix elements of $M^j_{AB}$ involving the holographic partners are automatically set to zero, so that \eqref{SymmetryForever} remains valid.}. Thus, we can conclude that $\varepsilon_A \varepsilon_B =-1$, when $M^j_{AB}\neq 0$. Hence, we are allowed to replace the second equation of \eqref{gabumbo} with
\begin{equation}\label{SymmetryForever}
M^j_{BA}= M^j_{AB} \, .
\end{equation} 

\subsection{Symmetric-Hyperbolicity and the birth of the Information Current}

If we introduce the notation $M^0_{AB}:= \mathcal{S}_{AB}$, the left-hand side of \eqref{LeFields} can be rewritten as $M^0_{AB}\partial_t \varphi^B$. This allows us to recast the field equations \eqref{LeFields} in the covariant Geroch-Lindblom form \cite{Geroch_Lindblom_1991_causal}:
\begin{equation}\label{Arrivati!}
M^\mu_{AB} \, \partial_\mu \varphi^B = -\Xi_{AB} \, \varphi^B \, .
\end{equation}
Now, we know from equation \eqref{SymmetryForever} that the matrices $M^j_{AB}$ are symmetric in $A$ and $B$. But also $M^0_{AB} (=\mathcal{S}_{AB})$ is symmetric, because it is the Hessian of the entropy. Additionally, $M^0_{AB}$ is positive definite, because the entropy has its absolute maximum at equilibrium. In conclusion, the system \eqref{Arrivati!} is symmetric hyperbolic. This is a surprising result: In the absence of macroscopic magnetic fields and spins, the OC symmetry principle is so strong that it constrains the mathematical structure of the field equations, enforcing symmetric-hyperbolicity. In some sense, Nature wants to give us the best equations possible! But there is more. Let us consider the quadratic vector field
\begin{equation}\label{nineth}
\mathbb{E}^\mu := \dfrac{1}{2} M^\mu_{AB} \varphi^A \varphi^B \,.
\end{equation}
If the hydrodynamic theory is causal, $\mathbb{E}^\mu$ is timelike future-directed \cite{Geroch_Lindblom_1991_causal}. Its four-divergence is $\partial_\mu \mathbb{E}^\mu = -\Xi_{AB}\varphi^A \varphi^B$, which is non-positive\footnote{The matrix $\Xi_{AB}$ is non-negative definite. This follows from the second law of thermodynamics. To see it, just compute the time-derivative of \eqref{DeltaS}  in the homogeneous limit: $\Delta \dot{S}= -V\varphi^A \dot{\mathbb{A}}_A = V \varphi^A \Xi_{AB} \varphi^B \geq 0$ (where $V$ is the volume occupied by the fluid). Also, notice that, if $M^\mu_{AB}$ were not symmetric, then, when we compute $\partial_\mu \mathbb{E}^\mu$, we would get an additional term $M^\mu_{[AB]}\varphi^B \partial_\mu \varphi^A$. Hence, equation \eqref{SymmetryForever} is crucial for the present considerations.}. Therefore, taken an arbitrary Cauchy surface $\Sigma$, the functional
\begin{equation}\label{Emumumu}
\mathbb{E} = \int_\Sigma \mathbb{E}^\mu d\Sigma_\mu \spc (\text{orientation: }d\Sigma_0 >0)
\end{equation}
is a non-increasing positive-definite norm of the perturbation. This guarantees \textit{covariant} stability of the theory \cite{Hishcock1983}. On the other hand, since $2\mathbb{E}^0 = \mathcal{S}_{AB}\varphi^A \varphi^B$, when we evaluate $\mathbb{E}$ on hypersurfaces $\{ t=\text{const} \}$, we find that it coincides with $-\Delta S$. Therefore, $\mathbb{E}^\mu$ can be interpreted as the information current of the system \cite{GavassinoGibbs2021,GavassinoCausality2021,
GavassinoStabilityCarter2022}. Note the generality of what we have just found: If the OC principle is respected, then we can always assign an information current to the system, even without postulating a priori the existence of an entropy current.

We can make one last observation. Suppose that we have a hydrodynamic theory whose entropy current $s^\mu$ is known, and it obeys the second law ($\partial_\mu s^\mu \geq 0$) as a strict mathematical inequality. Then, we can use the procedure outlined in \cite{GavassinoGibbs2021} to compute the information current. If the OC principle is respected (and there are no magnetic fields or spins), such information current should coincide with our current $\mathbb{E}^\mu$ \cite{GavassinoCausality2021}, defined in equation \eqref{nineth}. But, then, we can rewrite the system \eqref{Arrivati!} in an equivalent (manifestly symmetric) form:
\begin{equation}\label{EEEEE}
\partial_\mu \dfrac{\partial \mathbb{E}^\mu}{\partial \varphi^A} = -\Xi_{AB} \varphi^B \, .
\end{equation}
If the OC principle is respected, the principal part of the field equations is entirely determined by the information current! This result is similar to what we have found in a previous work \cite{GavassinoNonHydro2022}, but its interpretation is different. In \cite{GavassinoNonHydro2022}, equation \eqref{EEEEE} was just a procedure to construct ``nice theories''. Now, it constitutes an explicit technique to recast \textit{already existing theories} into a symmetric form. In fact, if the OC principle is respected, then the system \eqref{EEEEE} must be mathematically equivalent to the system of linearised field equations of the original theory, which usually is not presented in a symmetric form. Below, we give some examples, spanning different areas of relativistic hydrodynamics. 

\section{Some quick applications}

\subsection{Carter's multifluid theory}

Carter's multifluid theory is a formalism that enables us to model relativistic fluids with many chemical species, which are free to flow independently \cite{carter1991,Carter_starting_point,Andersson_2017,Termo}. It is renowned for its applications to Pulsar Glitch Theory \cite{langlois98,sourie_glitch2017,chamel_review_crust,
Geo2020,GavassinoIordanskii2021}, but it is receiving increasing attention because of its applicability to dissipation in relativistic superfluid systems \cite{Gusakov08,RauWasserman2020,GavassinoKhalatnikov2022}. Recently, we managed to prove that Carter's multifluid theory is causal and stable in the linear regime \cite{GavassinoStabilityCarter2022}. Is Carter's theory also symmetric-hyperbolic, in the linear regime?

The information current of Carter's theory has been computed in \cite{GavassinoStabilityCarter2022}. If we label the various chemical species of the multifluid using an abstract chemical index  $X$ (subject to Einstein's summation convention), then the components of the information current, as measured in the equilibrium rest frame, are 
\begin{equation}\label{tobedifferent}
\begin{split}
& T\mathbb{E}^0 = \dfrac{1}{2} \bigg[ \rho^{XY} \delta n_X \delta n_Y + \mathcal{K}^{XY} \delta j^k_X \delta j_{Yk} \bigg] \, , \\
& T \mathbb{E}^j = \rho^{XY} \delta j_X^j \delta n_Y \, , \\
\end{split}
\end{equation}
where $\delta n_X$ and $\delta j_X^k$ are linear perturbations to densities and currents (one for each chemical species $X$ of the system). The background matrices $\rho^{XY}$ and $\mathcal{K}^{XY}$ are both symmetric, and they are respectively the Hessian of the energy density (written as a function of the chemical densities), and the so-called ``entrainment matrix''. 

Let us apply the OC principle. Clearly, our array of perturbation fields is  $\varphi^A=\{ \delta n_X, \delta j_X^k \}$.
Since there are no magnetic fields or macroscopic spin tensors, and the degrees of freedom are ``standard'' (they are just densities and fluxes), equation \eqref{SymmetryForever} must hold. Thus, if Carter's theory is consistent with the OC principle, it should be possible to recast its field equations in the form \eqref{EEEEE}, namely (we multiply all equations by $T$ for convenience)
\begin{equation}\label{xyxyxyx}
\begin{split}
& \rho^{XY} \partial_t \delta n_Y + \rho^{XY} \partial_j \delta j_Y^j = B^{XY} \delta n_Y \, , \\
& \mathcal{K}^{XY}\partial_t \delta j_{Yk} + \rho^{XY}\partial_k \delta n_Y = C^{XY}\delta j_{Yk} \, , \\
\end{split}
\end{equation}
for some background matrices $B^{XY}$ and $C^{XY}$. This system of equations is manifestly symmetric hyperbolic, since both $\rho^{XY}$ and $\mathcal{K}^{XY}$ are positive definite (by stability \cite{GavassinoStabilityCarter2022}) and symmetric. Furthermore, it is straightforward to verify (compare with equation (84) of \cite{GavassinoStabilityCarter2022}) that the system \eqref{xyxyxyx} is indeed a rewriting of the linearised field equations of Carter's theory. Therefore, we have just proved that Carter's multifluid theory is symmetric-hyperbolic (in the linear regime), and it is consistent with the OC principle. This is good news for the neutron-star community.

\subsection{Israel-Stewart theory in the pressure frame}

As we anticipated in the introduction, the (linearised) Israel-Stewart theory is known to be symmetric-hyperbolic, both in the Eckart and in the Landau frame. However, there are other possible frames \cite{Bemfica2017TheFirst,NoronhaGeneralFrame2021}. For example, when modelling bulk viscosity, the Landau-frame prescription implements the viscous effects as corrections to the pressure \cite{Salmonson1991,Maartens1995,Camelio2022}. Instead, one may take an alternative approach, and implement bulk viscosity by adding a non-equilibrium correction $\mathcal{A}$ to the energy density \cite{Kovtun2019,BemficaDNDefinitivo2020,
GavassinoLyapunov_2020,DoreGavassino2022}. The resulting theory may be called ``Israel-Stewart theory in the  pressure frame'', because the equilibrium pressure matches the physical pressure. If we work (for simpliciy) at zero chemical potential, the stress-energy tensor and the entropy current of the Israel-Stewart theory in the pressure frame are
\begin{equation}\label{pressuro}
\begin{split}
& T^{\mu \nu}= (\rho+\mathcal{A}+P)u^\mu u^\nu +P g^{\mu \nu} \, , \\
& s^\mu = \bigg( s+\dfrac{\mathcal{A}}{T} - \dfrac{b_3 \mathcal{A}^2}{2T} \bigg) u^\mu \, , \\
\end{split}
\end{equation}
where $T$ is the temperature, $\rho(T)$ is the equilibrium energy density, $P(T)$ is the total pressure, and $s(T)$ is the equilibrium entropy density. These quantities are related by standard thermodynamic relations: $dP = s \, dT$, $d\rho=T \, ds$, and $\rho+P=Ts$. Additionally, we have $d\rho=c_v \, dT$, where $c_v(T)$ is the heat capacity per unit volume. The quantity $b_3(T)>0$ is a non-equilibrium transport coefficient. As field equations, we impose the energy-momentum conservation, $\partial_\mu T^{\mu \nu}=0$, and a formula for the entropy production: $T\partial_\mu s^\mu = \mathcal{A}^2/\lambda$, where $\lambda(T)>0$ is a transport coefficient. Then, it is easy to verify that the viscous degree of freedom $\mathcal{A}$ obeys the telegraph-type equation below:
\begin{equation}\label{linuozzoquiz}
\mathcal{A} = -\lambda \bigg[ \dfrac{u^\mu \partial_\mu T}{T} + b_3 u^\mu \partial_\mu \mathcal{A} + T\mathcal{A} \partial_\mu \bigg( \dfrac{b_3 u^\mu}{2T}\bigg)\bigg] \, .
\end{equation}
Is this theory symmetric-hyperbolic in the linear regime?

Let us apply the OC principle. Our ordered array of perturbation fields is $\varphi^A=\{\delta T, \delta u^k, \delta \mathcal{A} \}$. These fields constitute the degrees of freedom of the linearised theory. Again, there are no magnetic fields or macroscopic spins, and we have only  ``standard'' fields (densities and velocities), so that it should be possible to recast the linearised field equations in the form \eqref{EEEEE}. The information current can be easily computed from the constitutive relations \eqref{pressuro}, using the method outlined in \cite{GavassinoGibbs2021}. The result is
\begin{equation}\label{infurbo}
\begin{split}
& T\mathbb{E}^0 = \dfrac{1}{2} \bigg[ \dfrac{c_v (\delta T)^2}{T} + 2 \dfrac{\delta T \delta \mathcal{A}}{T} +  b_3 (\delta \mathcal{A})^2  + (\rho+P)\delta u^k \delta u_k \bigg] \, , \\
& T\mathbb{E}^j = s \, \delta T \delta u^j \, ,\\
\end{split}
\end{equation}
so that in our case equation \eqref{EEEEE} reads explicitly (again, multiplied by $T$)
\begin{equation}
\begin{split}
& T^{-1} \partial_t (c_v \delta T + \delta \mathcal{A}) + s \, \partial_j \delta u^j =0 \, , \\
& (\rho+P)\partial_t \delta u_k +s \, \partial_k \delta T =0  \, ,\\
& \partial_t ( b_3 \delta \mathcal{A} +T^{-1}\delta T ) = -\delta \mathcal{A}/\lambda  \, . \\
\end{split}
\end{equation}
Now, it is easy to verify that these are, indeed, the linearised field equations of the theory. The first one is the conservation of energy (divided by $T$), the second one is the conservation of momentum, and the third one is the telegraph-type equation \eqref{linuozzoquiz}. This system is presented in a manifestly symmetric form, and it is hyperbolic once we impose the rest-frame stability conditions ($\mathbb{E}^0 >0, \, \forall \, \varphi^A \neq 0$ \cite{GavassinoGibbs2021}). Hence, the Israel-Stewart theory in the pressure frame is symmetric-hyperbolic (in the linear regime), and it is consistent with the OC principle.

\subsection{Maxwell's equations in dispersionless isotropic media}

If there are magnetic fields, the reasoning that we followed to arrive at \eqref{SymmetryForever} is no longer valid. But that does not necessarily mean that equation \eqref{SymmetryForever} itself is not valid. We only need to be more careful about the couplings that the theory may present. Consider the example below.

Suppose that we want to model the propagation of (macroscopic) electromagnetic fields inside a medium, neglecting any possible macroscopic motion of matter. Then, our ordered array of degrees of freedom is just $\varphi^A=\{ E^k,H^k \}$, where $\textbf{E}$ is the electric field, and $\textbf{H}$ is the magnetic field \cite{jackson_classical_1999}. Now, the dynamics of $\textbf{H}$ can be coupled directly to that of $\textbf{E}$ through curl terms. However, $\textbf{E}$ is even under time reversal, whereas $\textbf{H}$ is odd. This implies that $\varepsilon_{\textbf{E}}\varepsilon_{\textbf{H}}=-1$, and equation \eqref{SymmetryForever} is still valid. Therefore, the macroscopic electromagnetic theory should still be symmetric hyperbolic, and it should still be possible to rewrite the field equations in the form \eqref{EEEEE}. Let's see if this is true.

First of all, we need a formula for the information current $\mathbb{E}^\mu$. We can derive it from simple thermodynamic reasoning. Assuming that the medium is dispersionless and isotropic, the energy associated to the presence of the electromagnetic field is (in the linear regime \cite{jackson_classical_1999})
\begin{equation}
U = \int \dfrac{1}{2} \, (\textbf{E}\cdot \textbf{D} + \textbf{B} \cdot \textbf{H}) \, d^3 x \, ,
\end{equation}
where $\textbf{D}=\epsilon \textbf{E}$ is the electric displacement, and $\textbf{B}=\mu \textbf{H}$ is the magnetic-flux density ($\epsilon$ and $\mu$ are background constants). We can assume that this energy is subtracted from the thermal energy $U_M$ of the medium, which may be treated as an ideal thermal bath, so that $U=-\Delta U_M = -T \Delta S$, where $T$ is the temperature of the medium. Therefore, recalling that $\mathbb{E}=-\Delta S$, we have that $U=T\mathbb{E}$, and we can identify $T\mathbb{E}^0$ with the electromagnetic energy density, and $T\mathbb{E}^j$ with the electromagnetic energy flux, which is just the Poynting vector, $\textbf{S}=\textbf{E}\times \textbf{H}$. In components,
\begin{equation}
\begin{split}
& T\mathbb{E}^0= \dfrac{1}{2} (\epsilon E^k E_k + \mu H^k H_k) \, , \\
& T\mathbb{E}^j= \epsilon\indices{^j _k _l} E^k H^l \, . \\
\end{split}
\end{equation}
If we use this information current to write the system \eqref{EEEEE} explicitly, setting the right-hand side to zero, we obtain (again, we multiply all equations by $T$)
\begin{equation}\label{gargone}
\begin{split}
& \partial_t \textbf{D}-\nabla \times \textbf{H}=0 \, ,\\
& \partial_t \textbf{B} + \nabla \times \textbf{E}=0 \, . \\
\end{split}
\end{equation}
These are just the dynamical Maxwell's equations in a medium, in the absence of macroscopic currents. Notice what we have found: a judicious definition for the entropy current, combined with the OC principle, has allowed us to derive the macroscopic Maxwell equations from purely thermodynamic reasoning. The system \eqref{gargone} is already expressed in a symmetric form. It may not look so, because of the ``$-$'' in the first equation, which does not appear in the second one. However, if we write explicitly the matrices $M^j_{AB}$, we immediately see that they are indeed symmetric:
\renewcommand{\arraystretch}{0.6}
\begin{equation}
TM^x_{AB}= 
\begin{bmatrix}
0 & 0 & 0 & 0 & 0 & 0  \\
0 & 0 & 0 & 0 & 0 & 1  \\
0 & 0 & 0 & 0 & -1 & 0  \\
0 & 0 & 0 & 0 & 0 & 0  \\
0 & 0 & -1 & 0 & 0 & 0  \\
0 & 1 & 0 & 0 & 0 & 0  \\
\end{bmatrix}  , \quad 
TM^y_{AB}= 
\begin{bmatrix}
0 & 0 & 0 & 0 & 0 & -1  \\
0 & 0 & 0 & 0 & 0 & 0  \\
0 & 0 & 0 & 1 & 0 & 0  \\
0 & 0 & 1 & 0 & 0 & 0  \\
0 & 0 & 0 & 0 & 0 & 0  \\
-1 & 0 & 0 & 0 & 0 & 0  \\
\end{bmatrix}  , \quad
TM^z_{AB}= 
\begin{bmatrix}
0 & 0 & 0 & 0 & 1 & 0  \\
0 & 0 & 0 & -1 & 0 & 0  \\
0 & 0 & 0 & 0 & 0 & 0  \\
0 & -1 & 0 & 0 & 0 & 0  \\
1 & 0 & 0 & 0 & 0 & 0  \\
0 & 0 & 0 & 0 & 0 & 0  \\
\end{bmatrix}  .
\end{equation}
Furthermore, implicit in the identification $U=-T\Delta S$, there was the assumption that $\epsilon>0$, and $\mu >0$ (since $\Delta S \leq 0$), so that $TM^0_{AB}=\text{diag}(\epsilon, \epsilon, \epsilon, \mu, \mu ,\mu)$ is positive definite, and the system \eqref{gargone} is symmetric-hyperbolic \cite{Volker2011}.

\subsection{Dealing with constraints: shear viscosity}

Suppose that we want to model the propagation of shear waves in a fluid. Then, the natural fields of our model are $\delta u^k$ (the perturbation to the flow velocity) and $\delta \Pi^{jk}$ (the perturbation to the shear stresses)\footnote{We do not need to include the pressure, or the density, among our degrees of freedom, because we are focusing on the pure ``shear sector'' (i.e. transverse waves), and no compression (or expansion) is allowed.}. 
Again, these are ``standard fluxes'', and the OC principle tells us that the resulting theory should be symmetric-hyperbolic. However, we face a technical subtlety here: the components of $\delta \Pi^{jk}$ are not all independent, because the shear stress tensor is symmetric and traceless: $\delta \Pi^{[jk]}=\delta \Pi \indices{^j _j}=0$. To be able to apply \eqref{EEEEE}, our fields $\varphi^A$ should instead constitute the algebraic degrees of freedom of the model. 
Luckily, there is an easy solution \cite{GavassinoNonHydro2022}: since the independent components of $\delta \Pi^{jk}$ are 5, we just need to introduce 5 thermodynamic degrees of freedom $Z^a$, $a=1,...,5$, which parameterize the state of the stress tensor. The constraints can be automatically enforced through the constitutive relation $\delta\Pi^{jk}=\delta\Pi^{jk}(Z^a)$. The simplest example of a constitutive relation that implements the constraints $\delta \Pi^{[jk]}=\delta \Pi \indices{^j _j}=0$ for any $Z^a$ is
\begin{equation}\label{akj}
\delta \Pi^{jk}(Z^a) =
\begin{bmatrix}
   Z^1+Z^2 & Z^3 & Z^4  \\
  Z^3 & Z^1-Z^2 & Z^5\\
  Z^4 & Z^5 & -2Z^1  \\
\end{bmatrix} = \dfrac{\partial \Pi^{jk}}{\partial Z^a} Z^a \, .
\end{equation} 
Clearly, our ordered array of degrees of freedom is $\varphi^A=\{ \delta u_k, Z^a \}$, which contains 8 independent field-variables. Applying \eqref{EEEEE} is now straightforward. The information current of the shear sector is well-known \cite{Hishcock1983,GavassinoGibbs2021,Almaalol2022}:
\begin{equation}
\begin{split}
& T\mathbb{E}^0= \dfrac{h}{2} \delta u^k \delta u_k + \dfrac{b_2}{2} \dfrac{\partial \Pi^{jk}}{\partial Z^a} \dfrac{\partial \Pi_{jk}}{\partial Z^b} Z^a Z^b \, , \\
& T\mathbb{E}^j= \dfrac{\partial \Pi^{jk}}{\partial Z^a} Z^a  \delta u_k  \, . \\
\end{split}
\end{equation}
Here, $h=\rho+P$ is the equilibrium enthalpy density, $b_2$ is a positive transport coefficient, and we used \eqref{akj} to express $\delta \Pi^{jk}$ in terms of $Z^a$. If we choose the right-hand side appropriately, the field equations that stem from \eqref{EEEEE} are
\begin{equation}\label{zab}
\begin{split}
 h \, \partial_t \delta u^k +\dfrac{\partial \Pi^{jk}}{\partial Z^a} \partial_j Z^a = {} & 0 \, ,\\
b_2 \dfrac{\partial \Pi^{jk}}{\partial Z^a} \dfrac{\partial \Pi_{jk}}{\partial Z^b} \partial_t Z^b +  \dfrac{\partial \Pi^{jk}}{\partial Z^a} \partial_j \delta u_k = {}& - \dfrac{\partial \Pi^{jk}}{\partial Z^a} \dfrac{\partial \Pi_{jk}}{\partial Z^b} \dfrac{Z^b}{2\eta} \, . \\
\end{split}
\end{equation} 
In this form, the equations look a bit strange. But let's take a closer look. The matrix $\partial \Pi^{jk}/\partial Z^a$ in the first equation can be brought inside the derivative, so that, with the aid of equation \eqref{akj}, we obtain $ h \, \partial_t \delta u^k +\partial_j \delta \Pi^{jk}=0$, which is just the conservation of momentum. We apply a similar procedure to the second equation, so that, if we multiply both sides by $2\eta$, we introduce the relaxation time $\tau=2\eta b_2$, and we bring all terms on the left-hand side, we obtain
\begin{equation}
 \dfrac{\partial \Pi^{jk}}{\partial Z^a} \bigg[ \tau \partial_t \delta \Pi_{jk} + \delta \Pi_{jk} + 2\eta \partial_j \delta u_k \bigg]=0 \, .
\end{equation}
This is just the Israel-Stewart relaxation equation for the shear stresses. In fact, the matrix $\partial \Pi^{jk}/\partial Z^a$ plays the only role of extracting the symmetric traceless part of $\partial_j \delta u_k$ (this can be easily verified explicitly). In conclusion, the system \eqref{zab} describes the shear sector of the Israel-Stewart theory, and it is a symmetric-hyperbolic system. Indeed, for the choice of constitutive relations \eqref{akj}, we have $TM^0_{AB}=\text{diag}(h,h,h,6b_2,2b_2,2b_2,2b_2,2b_2)$, with $h,b_2>0$, and\footnote{Note that, although \citet{Hishcock1983} showed that Israel-Stewart-like shear viscosity constitutes a symmetric-hyperbolic system, they did not give the explicit formulas for $M^y_{AB}$ and $M^z_{AB}$. With our method, it takes essentially no effort to compute them.}
\begin{equation}
TM^x_{AB}= 
\begin{bmatrix}
0 & 0 & 0 & 1 & 1 & 0 & 0 & 0  \\
0 & 0 & 0 & 0 & 0 & 1 & 0 & 0  \\
0 & 0 & 0 & 0 & 0 & 0 & 1 & 0  \\
1 & 0 & 0 & 0 & 0 & 0 & 0 & 0  \\
1 & 0 & 0 & 0 & 0 & 0 & 0 & 0  \\
0 & 1 & 0 & 0 & 0 & 0 & 0 & 0  \\
0 & 0 & 1 & 0 & 0 & 0 & 0 & 0  \\
0 & 0 & 0 & 0 & 0 & 0 & 0 & 0  \\
\end{bmatrix} ,
\, \,  
TM^y_{AB}= 
\begin{bmatrix}
0 & 0 & 0 & 0 & 0 & 1 & 0 & 0  \\
0 & 0 & 0 & 1 & -1 & 0 & 0 & 0  \\
0 & 0 & 0 & 0 & 0 & 0 & 0 & 1  \\
0 & 1 & 0 & 0 & 0 & 0 & 0 & 0  \\
0 & -1 & 0 & 0 & 0 & 0 & 0 & 0  \\
1 & 0 & 0 & 0 & 0 & 0 & 0 & 0  \\
0 & 0 & 0 & 0 & 0 & 0 & 0 & 0  \\
0 & 0 & 1 & 0 & 0 & 0 & 0 & 0  \\
\end{bmatrix} ,
\, \,
TM^z_{AB}= 
\begin{bmatrix}
0 & 0 & 0 & 0 & 0 & 0 & 1 & 0  \\
0 & 0 & 0 & 0 & 0 & 0 & 0 & 1  \\
0 & 0 & 0 & -2 & 0 & 0 & 0 & 0  \\
0 & 0 & -2 & 0 & 0 & 0 & 0 & 0  \\
0 & 0 & 0 & 0 & 0 & 0 & 0 & 0  \\
0 & 0 & 0 & 0 & 0 & 0 & 0 & 0  \\
1 & 0 & 0 & 0 & 0 & 0 & 0 & 0  \\
0 & 1 & 0 & 0 & 0 & 0 & 0 & 0  \\
\end{bmatrix} .
\end{equation}

\subsection{Connection with the GENERIC formalism}

The GENERIC formalism \cite{Grmela1997,Otting0_1998,
Otting2_1999,Otting3_1999,OttingerReview2018,OttingerReview2018} is a modern approach to non-equilibrium thermodynamics, where one assumes that the full dynamical evolution of a thermodynamic system can always be decomposed into two pieces: a Hamiltonian part (equipped with a Poisson-bracket structure), plus a purely irreversible part  (a sort of friction term). \citet{Pavelka2014} have shown that, if a theory has a GENERIC structure, then it is automatically consistent with the OC principle. On the other hand, here we have shown that, if there are no macroscopic magnetic fields or spins, then the equations of a relativistic hydrodynamic theory that is consistent with the OC principle can always be recast in the form \eqref{EEEEE}, in the linear regime. The implication is simple: in the absence of magnetic fields and spins, relativistic hydrodynamic theories based on the GENERIC formalism are a subset of the symmetric-hyperbolic theories presented in \cite{GavassinoNonHydro2022} (in the linear regime). This phenomenon has already been observed in \cite{GavassinoGENERIC2022}, for a specific GENERIC-based hydrodynamic theory \cite{Stricker2019}. Let us now explore this correspondence between the GENERIC formalism and the formalism of \cite{GavassinoNonHydro2022} more closely, with a more intuitive example.

We consider the most common fluid available: the perfect fluid at finite chemical potential. We choose its ordered array of degrees of freedom to be $\varphi^A=\{\delta \mu, \delta u^k, \delta T \}$, which describe the perturbations to respectively chemical potential, flow velocity, and temperature. The perfect-fluid information current \cite{GavassinoCausality2021}, expressed in the variables $\varphi^A$, is
\begin{equation}\label{tretisette}
\begin{split}
& T\mathbb{E}^0 = \dfrac{1}{2} \bigg[ \dfrac{\partial n}{\partial \mu}\bigg|_T (\delta \mu)^2+ 2\dfrac{\partial n}{\partial T}\bigg|_\mu \delta \mu \, \delta T + \dfrac{\partial s}{\partial T}\bigg|_\mu (\delta T)^2 + (\rho+P)\delta u^k \delta u_k \bigg] \, , \\
& T\mathbb{E}^j = n\, \delta \mu \delta u^j + s \, \delta T \delta u^j \, .\\
\end{split}
\end{equation}
To arrive at the formula above, one needs to invoke the thermodynamic identity $dP=nd\mu +sdT$, from which it is possible to derive the Maxwell relation \cite{landau5}
\begin{equation}\label{Maxwullo}
\dfrac{\partial n}{\partial T}\bigg|_\mu = \dfrac{\partial s}{\partial \mu}\bigg|_T \, .
\end{equation}
Given the information current \eqref{tretisette}, and assuming that $\Xi_{AB}=0$, the system \eqref{EEEEE}, multiplied by $T$, takes the form
\begin{equation}\label{perfucio}
\begin{split}
&  \dfrac{\partial n}{\partial \mu}\bigg|_T \partial_t \delta \mu + \dfrac{\partial n}{\partial T}\bigg|_\mu \partial_t \delta T + n \partial_j \delta u^j =0 \, , \\
& (\rho+P) \partial_t \delta u_k  + n \partial_k \delta \mu +s \partial_k \delta T =0 \, ,\\
& \dfrac{\partial s}{\partial T}\bigg|_\mu \partial_t \delta T + \dfrac{\partial n}{\partial T}\bigg|_\mu \partial_t \delta \mu + s \partial_j \delta u^j =0 \, .\\
\end{split}
\end{equation}
This system is manifestly symmetric. As consequence of stability, it is also symmetric-hyperbolic (we verify this explicitly in Appendix \ref{symmuzko}). Let us now analyse its equations one by one. In the first equation, we can combine together the time derivatives, and we obtain $\partial_t \delta n +\partial_j(n \delta u^j)=0$, which is just the continuity equation. If we combine together the space derivatives in the second equation, we obtain the conservation of momentum: $(\rho+P)\partial_t \delta u_k+\partial_k \delta P=0$. Finally, if we combine together the time derivatives in the third equation, and we invoke the Maxwell relation \eqref{Maxwullo}, we obtain the equation of entropy balance \cite{landau6}: $\partial_t \delta s+\partial_j (s\delta u^j)=0$. These are, indeed, the field equations of a perfect fluid. Now, let us recast \eqref{perfucio} in the form \eqref{LeFields}. Given the choice of fields $\varphi^A=\{ \delta \mu, \delta u^k, \delta T \}$, and the information current \eqref{tretisette}, it is easy to show that the conjugate fields are given by $T\mathbb{A}_A=\{\delta n, (\rho+P)\delta u_k , \delta s \}$. Thus, in order to recast \eqref{perfucio} in the form \eqref{LeFields} (rescaled by $T$), we only need to bring all space derivatives on the right-hand side:
\begin{equation}\label{confucianesimo}
\partial_t 
\begin{bmatrix}
\delta n \\[3pt]
(\rho+P)\delta u_x \\[3pt]
(\rho+P)\delta u_y \\[3pt]
(\rho+P)\delta u_z \\[3pt]
\delta s \\[3pt]
\end{bmatrix}
= -
\begin{bmatrix}
0 & n\, \partial_x & n\, \partial_y & n\, \partial_z & 0 \\[3pt]
n\, \partial_x & 0 & 0 & 0 & s\, \partial_x \\[3pt]
n\, \partial_y & 0 & 0 & 0 & s\, \partial_y \\[3pt]
n\, \partial_z & 0 & 0 & 0 & s\, \partial_z \\[3pt]
0 & s \, \partial_x & s\, \partial_y & s \, \partial_z & 0 \\[3pt]
\end{bmatrix}
\begin{bmatrix}
\delta \mu \\[3pt]
\delta u^x \\[3pt]
\delta u^y \\[3pt]
\delta u^z \\[3pt]
\delta T \\[3pt]
\end{bmatrix} \, .
\end{equation}
This system is already expressed in a GENERIC form. To see this, just compare with Section 2 of \citet{Otting1_1998}, making the following identifications: $T\mathbb{A}_A \rightarrow \text{``x''}$ (state variables), $-TM^{j}_{AB} \partial_j \rightarrow \text{``L''}$ (L-operator). In particular, note that the (homogeneous) equilibrium limit of equation (25) of \cite{Otting1_1998} coincides with the $5\times 5$ matrix in \eqref{confucianesimo}.

\subsection{IReD is the thermodynamic completion of DNMR}

We have applied the OC principle also to the DNMR theory in the Landau frame \cite{DMNR2012}. The analysis is essentially identical to that of section \ref{IS11}. There is, however, one subtle difference that is worth discussing. Within the DNMR theory, the diffusive current $\delta n_j$ and the shear stress tensor $\delta \pi_{jk}$ are dynamically coupled \cite{Sammet2023}:
\begin{equation}\label{margaret}
\begin{split}
 \tau_n \partial_t \delta n_j +\delta n_j ={}& ...+\ell_{n\pi}\partial_k \delta \pi^k_j , \\
\tau_\pi \partial_t \delta \pi_{jk} + \delta \pi_{jk} ={}&  ...-\ell_{\pi n} \braket{\partial_j \delta n_k} ,\\
\end{split}
\end{equation}
where $\braket{A_{jk}}$ is the symmetric traceless part of $A_{jk}$. The coupling coefficients $\ell_{n\pi}$ and $\ell_{\pi n}$ play in \eqref{margaret} a similar role as the role played by the Israel-Stewart coefficient $a_0$ in \eqref{astolfo}. Hence, just like the OC principle demands that $a_0$ be the same coefficient in the relaxation equations for both the bulk stress and the heat flux, similarly we find that the coefficients $\ell_{n\pi}$ and $\ell_{\pi n}$ must be related, for the OC principle to hold. In particular, the ratio $Z$ given below,
\begin{equation}
Z= -\dfrac{\kappa}{2\eta T} \dfrac{\ell_{\pi n}}{\ell_{n\pi}},
\end{equation}
must be equal to $1$ ($\eta$ is the shear viscosity and $\kappa$ is the charge diffusivity). Only under this condition, the DNMR theory is symmetric hyperbolic in the linear regime, and its equations of motion have a well posed initial value problem. It turns out that the value of $Z$ for a classical gas with constant cross section in the ultrarelativistic limit predicted by DNMR \cite{DMNR2012} is $0.29$. Thus, the OC principle is violated, and symmetric hyperbolicity is not longer guaranteed. 

When a violation of the OC principle occurs in a theory that arises from a truncation of a more complete theory (kinetic theory, in our case), this is a sign that the truncated theory is ``thermodynamically incomplete'', in the sense that some important terms have been left out \cite{GeigenmullerOnsagerApproximate1983}. In the case of DNMR, the problem comes from the second order terms in the Knudsen number (see $\mathcal{K}$, $\mathcal{K}^\mu$ and $\mathcal{K}^{\mu \nu}$ in \cite{DMNR2012}). These terms cannot be included in the hydrodynamic description, because they render the theory acausal (and hence unstable \cite{GavassinoSuperluminal2021}), but they can neither be ignored in our analysis, since some of them survive in the linear regime (e.g. the term proportional to $\partial^k \braket{\partial_k \delta u_j}$). However, it has been recently shown that, by appropriately redefining the matching scheme, it is possible to ``reabsorb'' all these problematic terms into the definition of the transport coefficients \cite{Wagner2022}. The result is the IReD theory, whose fluid equations are formally equivalent to those of DNMR, but with different values of the transport coefficients (and with $\mathcal{K}=\mathcal{K}^\mu=\mathcal{K}^{\mu \nu} \equiv 0$). As one would expect, in IReD, when the contributions from all the moments are included, $Z$ is identically $1$ (use the values from Tables III and IV of \cite{Wagner2022}). We can thus conclude that IReD is the ``thermodynamic completion'' of DNMR. In IReD, symmetric hyperbolicity and consistency with the OC principle are restored.

\section{Conclusions}

Usually, one invokes the Onsager-Casimir (OC) principle to constrain the transport coefficients of a dissipative fluid model (e.g. the bulk viscosities of a superfluid \cite{landau6}, or the rates of a reacting mixture \cite{carter1991}). Here, we have invoked the OC principle to set constraints on the mathematical structure  of the hydrodynamic theory as a whole. To achieve this goal, we have expanded the hydrodynamic fields in the sine-cosine Fourier basis, and we have treated the linear combination coefficients as non-equilibrium thermodynamic variables, whose dynamics should be consistent with the OC principle. The result is surprisingly simple: if the degrees of freedom of are the ``usual'' densities and fluxes, then the hydrodynamic theory must necessarily be symmetric-hyperbolic.

We have also formulated a simple technique that allows one to recast the linearised field equations of a hydrodynamic theory in a manifestly symmetric form, provided that the theory is consistent with the Onsager-Casimir principle. This technique just involves computing the derivatives of the information current $\mathbb{E}^\mu$ \cite{GavassinoCausality2021} with respect to the components of the fields [see equation \eqref{EEEEE}]. One of the implications is that, if the Onsager-Casimir principle is obeyed, then the principal part of the hydrodynamic equations (the part with highest derivatives) can be uniquely determined from thermodynamic considerations alone. We have applied these ideas to Carter's multifluid theory, to the Israel-Stewart theory (for bulk viscosity) in the pressure frame, to Maxwell's equations in dispersionless isotropic media, and to the shear sector of the Israel-Stewart theory. It worked. In all four examples, we managed to recast the field equations in a manifestly symmetric form, proving that such theories are symmetric-hyperbolic (for Maxwell's equations, this was already known \cite{Volker2011}). This same methodology had already been successfully employed to recast the ``GENERIC theory'' \cite{GavassinoGENERIC2022} in a symmetric form, although the thermodynamic foundations of the procedure were not clear, yet. Indeed, one of the implications of our work is that, in the linear regime, any relativistic hydrodynamic theory that is consistent with the GENERIC formalism is symmetric-hyperbolic, if the fields are ``standard''. 

Finally, we have verified that the DNMR theory \cite{DMNR2012} violates the Onsager-Casmir principle. This reflects the ``thermodynamic incompleteness'' of the DNMR truncation, as shown by the necessity of neglecting the second order terms in the Knudsen number, which would render the field equations unstable. Consistency with the Onsager-Casmir principle is however restored within the IReD framework \cite{Wagner2022}, which may be regarded as the ``thermodynamic completion'' of DNMR.

So, what's next? First of all, this work has shown that the procedure for constructing relativistic hydrodynamic theories outlined in \cite{GavassinoNonHydro2022} has solid thermodynamic foundations. It is the most rigorous way of constructing linear theories that are consistent with the Onsager-Casimir principle. Secondly, this work has also revealed that there may be physically-motivated exceptions to symmetric-hyperbolicity, if the hydrodynamic fields are ``exotic'' enough. Everything depends on their behaviour under time reversal. So, what if we add exotic degrees of freedom that may indeed lead to a breakdown of symmetric-hyperbolicity? For example, suppose that we include in our equations some additional fields such that, when we apply the Onsager-Casimir principle, the resulting theory is necessarily elliptic. This would surely constitute a violation of causality \cite{Rauch_book}. Hence, these theories should probably be rejected. Does this mean that we can combine the OC principle with the principle of causality to determine which macroscopic fields are allowed to exist, and which are not? This may be good material for future investigations.

\section*{Acknowledgements}

This work was supported by a Vanderbilt's Seeding Success Grant. I thank M. Disconzi, J. Noronha, M. Antonelli, G. Torrieri and D. Montenegro for reading the manuscript and providing useful comments. I also would like to thank D. Rischke, D. Wagner, and V. Ambru\textcommabelow{s} for  an illuminating discussion about the connection between DNMR, IReD, and the OC principle. Finally, I am grateful to the anonymous referee, whose guidelines helped me improve the clarity of the presentation.

\appendix

\section{Different paths to the Onsager-Casmir relations}


\subsection{Matrix manipulations}\label{camusso}

Our set of independent variables is the $8\mathfrak{D}$-array $\{ \mathbb{A}^1_A,\mathbb{A}^2_A,...,\mathbb{A}^8_A \}$, whose conjugate array is $\{\varphi_1^A,\varphi_2^A,...,\varphi_8^A \}$. If we introduce the $\mathfrak{D}\times\mathfrak{D}$ matrices $\Xi=[\Xi_{AB}]$, $M^j=[k_j M^j_{AB}]$ (no sum over $j$), and $\varepsilon=\text{diag}[\varepsilon_A]$, then the matrix of kinetic coefficients, denoted by \citet{Onsager_Casimir} as ``$p$'', and the time-reversal matrix $\mathcal{E}$ \cite{Krommes1993} are respectively 
\begin{equation}
p = 
\begin{bmatrix}
\Xi & - M^x & - M^y & - M^z & 0 & 0 & 0 & 0\\
 M^x & \Xi  & 0 & 0 & - M^y & - M^z & 0 & 0\\
 M^y & 0 & \Xi  & 0 & - M^x & 0 & - M^z & 0\\
 M^z & 0 & 0 & \Xi  & 0 & - M^x & - M^y & 0\\
0 &  M^y & M^x & 0 & \Xi  & 0 & 0 & - M^z \\
0 &  M^z & 0 &  M^x & 0 & \Xi  & 0 & - M^y\\
0 & 0 &  M^z &  M^y & 0 & 0 & \Xi  & - M^x\\
0 & 0 & 0 & 0 &  M^z &  M^y &  M^x & \Xi \\
\end{bmatrix}, \quad \quad 
\mathcal{E}=
\begin{bmatrix}
\varepsilon & 0 & 0 & 0 & 0 & 0 & 0 & 0 \\
0 & \varepsilon & 0 & 0 & 0 & 0 & 0 & 0 \\
0 & 0 & \varepsilon & 0 & 0 & 0 & 0 & 0 \\
0 & 0 & 0 & \varepsilon & 0 & 0 & 0 & 0 \\
0 & 0 & 0 & 0 & \varepsilon & 0 & 0 & 0 \\
0 & 0 & 0 & 0 & 0 & \varepsilon & 0 & 0 \\
0 & 0 & 0 & 0 & 0 & 0 & \varepsilon & 0 \\
0 & 0 & 0 & 0 & 0 & 0 & 0 & \varepsilon \\
\end{bmatrix} \, .
\end{equation}
The OC symmetry principle states that $p=\mathcal{E}p^\dagger\mathcal{E}^\dagger$ \cite{Krommes1993}. On the other hand, we can write $\mathcal{E}p^\dagger\mathcal{E}^\dagger$ explicitly:
\begin{equation}
\mathcal{E}p^\dagger\mathcal{E}^\dagger = 
\begin{bmatrix}
\varepsilon \Xi^\dagger \varepsilon^\dagger &  \varepsilon M^{x\dagger}\varepsilon^\dagger & \varepsilon M^{y\dagger}\varepsilon^\dagger & \varepsilon M^{z\dagger}\varepsilon^\dagger & 0 & 0 & 0 & 0\\
-\varepsilon M^{x\dagger} \varepsilon^\dagger & \varepsilon \Xi^\dagger \varepsilon^\dagger  & 0 & 0 & \varepsilon M^{y\dagger} \varepsilon^\dagger & \varepsilon M^{z\dagger} \varepsilon^\dagger & 0 & 0\\
-\varepsilon M^{y\dagger} \varepsilon^\dagger & 0 & \varepsilon \Xi^\dagger \varepsilon^\dagger  & 0 & \varepsilon M^{x\dagger} \varepsilon^\dagger & 0 & \varepsilon M^{z\dagger} \varepsilon^\dagger & 0\\
-\varepsilon M^{z\dagger} \varepsilon^\dagger & 0 & 0 & \varepsilon \Xi^\dagger \varepsilon^\dagger  & 0 & \varepsilon M^{x\dagger} \varepsilon^\dagger &  \varepsilon M^{y\dagger} \varepsilon^\dagger & 0\\
0 & -\varepsilon M^{y\dagger} \varepsilon^\dagger & - \varepsilon M^{x\dagger} \varepsilon^\dagger & 0 & \varepsilon \Xi^\dagger \varepsilon^\dagger  & 0 & 0 & \varepsilon M^{z\dagger} \varepsilon^\dagger \\
0 & -\varepsilon M^{z\dagger} \varepsilon^\dagger & 0 & -\varepsilon M^{x\dagger} \varepsilon^\dagger & 0 & \varepsilon \Xi^\dagger \varepsilon^\dagger  & 0 &  \varepsilon M^{y\dagger} \varepsilon^\dagger \\
0 & 0 & -\varepsilon M^{z\dagger} \varepsilon^\dagger & -\varepsilon M^{y\dagger} \varepsilon^\dagger & 0 & 0 & \varepsilon \Xi^\dagger \varepsilon^\dagger  & \varepsilon M^{x\dagger} \varepsilon^\dagger\\
0 & 0 & 0 & 0 & -\varepsilon M^{z\dagger} \varepsilon^\dagger &  -\varepsilon M^{y\dagger} \varepsilon^\dagger & -\varepsilon M^{x\dagger} \varepsilon^\dagger & \varepsilon \Xi^\dagger \varepsilon^\dagger \\
\end{bmatrix} \, .
\end{equation}
It is immediate to see that $p=\mathcal{E}p^\dagger\mathcal{E}^\dagger$ holds if and only if $\Xi=\varepsilon \Xi^\dagger \varepsilon^\dagger$ and  $ M^j=-\varepsilon M^{j\dagger} \varepsilon^\dagger$, which is just equation \eqref{gabumbo} expressed using matrix notation.

\subsection{Retracing Casimir's reasoning}\label{AAA}

In this appendix, we show how to generalise the reasoning of \citet{Onsager_Casimir} to a hydrodynamic setting. 

First of all, let us recall that the microcanonical probability distribution is $\mathcal{P}\propto e^{\Delta S}$. Considering that, in our case, the degrees of freedom are fields, all microcanonical averages are functional integrals:
\begin{equation}
\braket{f} = \dfrac{\int \mathcal{D} \varphi^A \,  e^{\Delta S[\varphi^A]} f[\varphi^A]}{\int \mathcal{D} \varphi^A \, e^{\Delta S[\varphi^A]} } \, ,
\end{equation}
for any $f=f[\varphi^A]$.
If we approximate $\Delta S$ using equation \eqref{DeltaS}, then the functional integral is Gaussian, and we have the well-known relation
\begin{equation}\label{deltaxx}
\braket{\mathbb{A}_A(0,\textbf{x}) \varphi^B(0,\tilde{\textbf{x}})}= \delta \indices{^B _A} \delta^3(\textbf{x}-\tilde{\textbf{x}}) \, .
\end{equation}
This formula is the analogue of equation \eqref{ConjugoA} for fields. It should be noted that equation \eqref{deltaxx} cannot be interpreted too literally, because the Dirac delta gives equal weight to all wavenumbers, and we know that, when $\textbf{k}$ becomes large, the gradient corrections in \eqref{DeltaS} become relevant. However, since we are only interested in the behaviour of the system at long wavelengths, the error that we commit in equation \eqref{deltaxx} will not affect our final result.

To derive the OC symmetry relations, one starts from the equation of microscopic reversibility \cite{Geigenmuller1983,Carbone2020}:
\begin{equation}
\braket{\mathbb{A}_A(t,\textbf{x}) \mathbb{A}_B(0,\tilde{\textbf{x}})} = \braket{\mathbb{A}_A(0,\textbf{x}) \mathbb{A}_B(t,\tilde{\textbf{x}})} \varepsilon_A \varepsilon_B \, ,
\end{equation}
which expresses the symmetry of the equilibrium fluctuations under time reversal \cite{landau5}. If we take its (macroscopic \cite{Onsager_Casimir}) time-derivative at $t=0$, and we invoke Onsager's ``regression hypothesis'' \cite{Geigenmuller1983}, i.e. we assume that the correlation functions are governed by the macroscopic field equations \eqref{LeFields}, we obtain
\begin{equation}
\bigg\langle \bigg(\Xi_{AC}  +M^j_{AC} \dfrac{\partial}{\partial x^j} \bigg) \varphi^C(0,\textbf{x}) \, \mathbb{A}_B(0,\tilde{\textbf{x}}) \bigg\rangle = \bigg\langle\mathbb{A}_A(0,\textbf{x}) \bigg(\Xi_{BC} +M^j_{BC} \dfrac{\partial}{\partial \tilde{x}^j} \bigg) \varphi^C(0,\tilde{\textbf{x}})\bigg\rangle \varepsilon_A \varepsilon_B \, .
\end{equation}
Recalling equation \eqref{deltaxx}, we can rewrite the above identity in terms of Dirac deltas:
\begin{equation}
\Xi_{AB} \delta^3(\textbf{x}-\tilde{\textbf{x}}) + M^j_{AB} \dfrac{\partial}{\partial x^j} \delta^3(\textbf{x}-\tilde{\textbf{x}}) = \Xi_{BA} \varepsilon_A \varepsilon_B \delta^3(\textbf{x}-\tilde{\textbf{x}}) + M^j_{BA} \varepsilon_A \varepsilon_B   \dfrac{\partial}{\partial \tilde{x}^j}\delta^3(\textbf{x}-\tilde{\textbf{x}}) \, .
\end{equation}
Equation \eqref{gabumbo} follows\footnote{Use the general fact that $\partial_a \delta(a-b)=-\partial_b \delta(a-b)$.}.

Indeed, we would like to mention that there are previous instances in the literature where the OC principle has been applied to a field theory as a whole. In particular, consider the approach of \citet{Hubmer1988} (Section 2). The idea is the following: if the field equations are expressed in the form $\partial_t \mathbb{A}_A =-\mathcal{L}_{AB}\varphi^B$, where $\mathcal{L}_{AB}$ are operators on $L^2(\mathbb{R}^3)$, and the fields $\mathbb{A}_A$ and $\varphi^A$ are related (as in our case) by the duality condition
\begin{equation}
\varphi^A(\textbf{x}) = -\dfrac{\delta (\Delta S)}{\delta \mathbb{A}_A(\textbf{x})} \, ,
\end{equation}
then the operators $\mathcal{L}_{AB}$ satisfy the symmetry condition $\mathcal{L}_{BA}=\varepsilon_A \varepsilon_B \mathcal{L}_{AB}^\dagger$, where ``$\dagger$'' denotes the standard Hermitian conjugation of operators on $L^2(\mathbb{R}^3)$. It is clear that equation \eqref{LeFields} takes indeed the form considered in \cite{Hubmer1988}, with $\mathcal{L}_{AB}=\Xi_{AB}+M^j_{AB}\partial_j$. Considering that $\partial_j^\dagger=-\partial_j$, equation \eqref{gabumbo} is again recovered.

\section{Symmetric-hyperbolicity of perfect fluids}\label{symmuzko}

The stability conditions of a perfect fluid are $\rho+P>0$ (positive inertia \cite{GavassinoStabilityCarter2022,MTW_book}), $c_p>0$ (stability against thermal fluctuations; $c_p$ is the specific heat at constant pressure), and $0<c_s^2\leq 1$ (stability against compression and causality \cite{GavassinoCausality2021,GavassinoSuperluminal2021}). \citet{Hishcock1983} have shown that many other thermodynamic inequalities follow directly from these conditions. For example, with the aid some Maxwell relations, one can derive the well-known inequalities
\begin{equation}\label{cavazzup}
\begin{split}
 c_v >{}& 0 \, , \\
 \dfrac{\partial n}{\partial (\mu/T)}\bigg|_T >{}& 0 \, , \\
\end{split}
\end{equation}
see equations (96) and (101) of \cite{Hishcock1983}. Our goal here is to prove that the system \eqref{perfucio} is symmetric-hyperbolic, and this follows from the thermodynamic inequalities above. In fact, since the system \eqref{perfucio} is clearly symmetric, we only need to prove that the matrix $M^0_{AB}$ is positive-definite. Given that $\rho+P>0$, we just need to show that the $2\times 2$ block
\begin{equation}
\begin{bmatrix}
   \dfrac{\partial n}{\partial \mu}\bigg|_T & \dfrac{\partial n}{\partial T}\bigg|_\mu   \\
  \dfrac{\partial n}{\partial T}\bigg|_\mu &  \dfrac{\partial s}{\partial T}\bigg|_\mu \\
\end{bmatrix}
\end{equation} 
is itself positive-definite. However, because of the second inequality in \eqref{cavazzup}, we know that the first diagonal element is positive. Thus, if we manage prove that the determinant is positive, we are done. To this end, a well-known thermodynamic identity comes to our aid (see \citet{landau5}, \S 24 problem 1):
\begin{equation}
\dfrac{\partial n}{\partial \mu}\bigg|_T \, \dfrac{\partial s}{\partial T}\bigg|_\mu -\bigg(\dfrac{\partial n}{\partial T}\bigg|_\mu\bigg)^2 = \dfrac{c_v}{T} \, \dfrac{\partial n}{\partial \mu}\bigg|_T  > 0\, ,
\end{equation}
where the inequality on the right is a consequence of \eqref{cavazzup}. This completes our proof.

\bibliography{Biblio}

\label{lastpage}

\end{document}